\documentclass[11pt]{article}

\usepackage{amsmath,amssymb,latexsym}
\usepackage[shortcuts,cyremdash]{extdash}
\usepackage{graphics,ifthen,epsfig}
\usepackage{xspace}
\usepackage{relsize}
\usepackage{afterpage}
\usepackage{setspace}
\usepackage[hang,small,bf]{caption}
\usepackage{natbib}
\usepackage{subfigure}
\usepackage{endnotes,etoolbox}
\usepackage{float}
\usepackage[hang,flushmargin]{footmisc} 
\usepackage[top=1in, bottom=1in, left=1in, right=1in]{geometry}
\usepackage{color}
\usepackage{bbm}

\bibpunct{(}{)}{;}{a}{}{,}

\newcommand{\betastar}{\beta^*}
\newcommand{\betainf}{\underline{\beta}}

\newcommand{\ncm}[2]{\newcommand{#1}{\ensuremath{#2}\xspace}}

\ncm{\fa}{A}

\newtheorem{theorem}{Theorem}
\newtheorem{proposition}{Proposition}
\newtheorem{corollary}{Corollary}
\newtheorem{lemma}{Lemma}

\newlength{\tempind}
\newlength{\tempskip}

\newenvironment{proof}
{ \setlength{\tempind}{\parindent} \setlength{\tempskip}{\parskip}
\setlength{\parindent}{0.25in} \setlength{\parskip}{2 pt} \noindent
\textbf{Proof.} }
{%
\setlength{\parindent}{\tempind} \setlength{\parskip}{\tempskip}
\hfill $\blacksquare$ \vspace{2mm} \\}

\let\OLDthebibliography\thebibliography
\renewcommand\thebibliography[1]{
	\OLDthebibliography{#1}
	\setlength{\parskip}{0pt}
	\setlength{\itemsep}{0pt}
	\vspace*{-12pt}
}

\begin{document}

\begin{center}
{\LARGE A Diffusion Model of Dynamic Participant Inflow Management }\\[12pt]
{\small Baris Ata, Booth School of Business, University of Chicago,
5807 S. Woodlawn Ave, Chicago, IL 60637, USA, baris.ata@chicagobooth.edu, p:773-834-2344, ORCID:0009-0001-4793-4607} \\ [3pt]
{\small Deishin Lee, Ivey Business School at Western University, 1255 Western Rd, London, ON N6G0N1, Canada, dlee@ivey.ca, p:519-661-3288, ORCID:0000-0002-6335-2877}  \\ [3pt]
{\small Mustafa H. Tongarlak, Bo\u{g}azi\c{c}i University, Bebek, 34342 Be\c{s}ikta\c{s}/Istanbul, T\"{u}rkiye, tongarlak@boun.edu.tr, p:+90 212 359 6503, ORCID:0000-0002-4812-2539} \\ [6pt]
{\small October 30, 2023} \\
\end{center}

\doublespacing
\begin{abstract}

\noindent This paper studies a diffusion control problem motivated by challenges faced by public health agencies who run clinics to serve the public. A key challenge for these agencies is to motivate individuals to participate in the services provided. They must manage the flow of (voluntary) participants so that the clinic capacity is highly utilized, but not overwhelmed. The organization can deploy costly promotion activities to increase the inflow of participants. Ideally, the system manager would like to have enough participants waiting in a queue to serve as many individuals as possible and efficiently use clinic capacity. However, if too many participants sign up, resulting in a long wait, participants may become irritated and hesitate to participate again in the future. We develop a diffusion model of managing participant inflow mechanisms. Each mechanism corresponds to choosing a particular drift rate parameter for the diffusion model. The system manager seeks to balance three different costs optimally: i) a linear holding cost that captures the congestion concerns; ii) an idleness penalty corresponding to wasted clinic capacity and negative impact on public health, and iii) costs of promotion activities. We show that a nested-threshold policy for deployment of participant inflow mechanisms is optimal under the long-run average cost criterion. In this policy, the system manager progressively deploys mechanisms in increasing order of cost, as the number of participants in the queue decreases. We derive explicit formulas for the queue length thresholds that trigger each promotion activity, providing the system manager with guidance on when to use each mechanism.

\end{abstract}

\newpage
\doublespacing
\abovedisplayskip=9pt
\abovedisplayshortskip=0pt
\belowdisplayskip=9pt
\belowdisplayshortskip=9pt

%%%%%%%%%%%%%%%%%%%%%%%%%%
% Section: Introduction
%%%%%%%%%%%%%%%%%%%%%%%%%%

\section{Introduction} \label{sec:intro}

This paper studies a diffusion control problem motivated by challenges faced by public health agencies who run clinics to serve the public. Examples include mass vaccination clinics and blood donation clinics. A key challenge for these agencies tasked with serving the public is to motivate individuals to participate in the services provided. Public health officials cannot control whether and when individuals arrive at the clinics. They can use promotion activities to increase public awareness and thus the participant inflow at clinics, but deploying such mechanisms can be costly. These activities can range from simple email communications and public health advertisements to speaking events for public health officers. If participation turnout is low, there is a negative impact on public health and wasted clinic capacity.

We use a diffusion model to analyze such a service organization where a system manager can deploy costly participant inflow mechanisms to increase the rate at which participants sign up.  Ideally, the system manager would like to have enough participants waiting in a queue to serve as many individuals as possible and efficiently use clinic capacity. However, if too many participants sign up, resulting in a long wait, participants may become irritated and hesitate to sign up again in the future (e.g., for the next vaccine dose or giving blood in the future). This is captured by a holding cost in our model. The system manager must use these promotion activities judiciously to balance the inflow of participants, with the risk of overspending on promotion if too many participants sign up.

We show that a nested threshold policy is optimal for the deployment of promotion activities. In this policy, the system manager should progressively deploy these mechanisms in increasing order of cost, as the number of participants in the queue decreases. We derive the explicit queue length thresholds that trigger each promotion activity, providing the system manager with guidance on when to use each mechanism.

The methodological contribution of the paper lies in its solution of a drift-rate control problem on an unbounded domain with state costs. The associated Bellman equation involves boundary conditions at the origin and at infinity. The former is a Neumann type boundary condition whereas the latter is a linear growth condition for the derivative of the value function. Our solution approach proceeds by solving a family of initial value problems starting at the origin parameterized by the average cost rate $\beta$ (defined below), initially ignoring the aforementioned growth condition. We develop an approach to pin down the unique value of the average cost rate $\beta$ that ensures the linear growth condition, which is the technical novelty of the paper.

\paragraph{Literature Review.}   

Our problem can be viewed as controlling the arrival rate of a queueing system. Such problems have often been tackled by either heavy traffic approximations or Markov decision process formulations in the literature. The heavy traffic approach, pioneered by \citet{Harrison_88}, approximates the original control problem for a queueing system with a diffusion control problem, which is easier to analyze; see \citet{Harrison_Wein_89,Harrison_Wein_90} for early examples of this approach. For problems similar to ours, heavy traffic approximations often result in drift rate control problems. A closely related paper to ours is \citet{Ata_Drift_Control}, which considers a drift rate control problem on a bounded interval under a general cost of control but no holding costs. \citet{Ata_Thin_Arrival_Streams} builds on \citet{Ata_Drift_Control} and approximates a multi-class make-to-order production system with a drift rate control problem on a bounded interval with a piecewise linear convex cost of control. Methodologically, our paper relates to \citet{Ata_Thin_Arrival_Streams} but it extends it in two important ways. First, its state space is unbounded. Second, it incorporates holding costs. Incorporating these two important model features leads to a significantly more complex analysis. Another closely related paper is \citet{ata-etal-2023}. This paper differs from \citet{ata-etal-2023} in that the participants (in the public health setting) are unlikely to abandon. Thus, we do not incorporate abandonments in our model. This modeling difference leads to a different type of Bellman equation, namely, one that involves a growth condition; and its solution requires a different approach.

\citet{Ghosh_2007} extends \citet{Ata_Drift_Control} by incorporating holding costs and allowing the system manager to choose the bounded interval on which the process lives endogenously. \citet{Ghosh_2010}  extends this work by introducing abandonments; also see \citet{Ata_MakeToOrder_2009}, \citet{Ghamami_Ward_2013}, \citet{Ata_Tongarlak_Queueing_2013}, and \citet{Sun_2020} for similar formulations with abandonments. In other related work, \citet{ata-etal-2019} approximates a gleaning operation using a drift rate control problem, and derives a nested threshold policy as an optimal staffing policy. The authors derive an approximation in the many server asymptotic regime. The authors derive an approximation in the many server asymptotic regime. Consequently, the drift rate of their control problem has a different structure. This makes their analysis inapplicable in our context.

Recent work by \citet{ata-barjesteh-2019} studies dynamic pricing, scheduling and outsourcing decision for a make-to-stock manufacturing system. They approximate this problem by a drift-rate control problem with a quadratic cost of control. Leveraging the elegant structure of the Riccati equation that arises as the Bellman equation, the authors derived a closed form solution for the optimal dynamic prices in terms of Airy functions. Several researchers established the asymptotic optimality of their proposed policies in the heavy traffic limit. For example, \citet{Budhiraja_2011} studies the service rate and admission control for a queueing network and derives policies that are asymptotically optimal; also see \citet{Bell_Williams_2001, bell-williams-2005}, \citet{ata-kumar-2005}, \citet{Ata_Olsen_2009, Ata_Olsen_Queueing_2013} for asymptotically optimal policies in other related settings.

Many researchers have used Markov decision process formulations to study admission control or service rate problems; see for example \citet{Crabill_72, Crabill_74}. \citet{Stidham_Weber_89} studies monotone control policies for dynamic control of service and arrival rates to a queueing network. \citet{George_Harrison_2001} considers the dynamic service rate control problem for an M/M/1 queue and derives the optimal policy essentially in closed form. 
Similarly, \citet{Ata_Shneorson_2006} considers a dynamic arrival and service rate control problem for an M/M/1 queue.
\citet{Ata_2005} and \citet{Ata_Zachariadis_2007} consider related service rate control problems for queueing systems arising in wireless communications applications and solve them explicitly; also see \citet{Hasenbein_2010} and \citet{Lewis_2013} for studies of other related questions.

%%%%%%%%%%%%%%%%%%%%%%%%%%%%%%
% MODEL
%%%%%%%%%%%%%%%%%%%%%%%%%%%%%%

\section{Model} \label{sec:model}

All stochastic processes live on a filtered probability space $(\Omega, \mathcal{F}, \mathbb{P}; \mathcal{F}_t, t \ge 0)$ that satisfies the usual conditions; see \citet{harrison-2013}.
In the absence of control, we model the participant queue as a reflected Brownian motion with drift parameter $\theta_0 < 0$ and variance parameter $\sigma^2 < \infty$. At any point in time, the system manager can engage in activities $1, \dots , K$. Activity $k$ increases the drift rate by $\mu_k > 0$, (i.e. it increases the participant arrival rate) but costs $c_k \mu_k$ per unit of time. We assume $ 0 < c_1 < ... < c_K$,
 %\label{eqn:ck} %
and model the system manager's control by a K-dimensional process  $\delta = (\delta_k)$, where
\begin{align}
\delta_k(t) = \Bigg\{ \begin{array}{ll}
     1 & \ \ \ \mbox{if the system manager engages in activity}\, k,  \\
     0 & \ \ \ \mbox{otherwise.}
     \end{array}
  \label{eqn:deltat}
\end{align}
The system manager's problem can then be expressed as follows: Choose $\delta = (\delta_k)$ adapted to $\{\mathcal{F}_t, t \ge 0\}$ so as to 
\begin{align}
& \min \limsup_{t \rightarrow \infty} \frac{1}{t} \mathbb{E} \left[
      \int_{0}^{t} \sum_{k=1}^{K}  c_k \mu_k \delta_k(s) ds
     + h \int_{0}^{t}  Z(s) ds + p L(t)
     \right] \label{eqn:appox-obj2} \\ 
& \mbox{such that} \nonumber \\
&  Z(t) = X(t) + \theta_0 t + \int_{0}^{t} \sum_{k=1}^{K}  \mu_k \delta_k(s) ds + L(t), \,\, t \ge 0, \label{eqn:approx-cons2.1} \\
&  \,\,  Z(t) \ge 0, \,\, t \ge 0, \label{eqn:approx-cons2.2} \\
&  \,\,  \delta_k(t) \in [0,1], \,\, t \ge 0, \,\, k = 1, ... , K, \label{eqn:approx-cons2.3} \\
&  \,\, \int_0^{\infty} 1_{\{Z(s) > 0\}} dL(s) = 0,  \label{eqn:approx-cons2.4} \\
&  \,\, L, \delta_k(\cdot), \,\, k = 1, ... K, \mbox{ adapted to } \{\mathcal{F}_t, t \ge 0\}, \label{eqn:approx-cons2.5} 
\end{align}
where $X$ is a $(0, \sigma^2)$ Brownian motion with $X(0) = 0$ almost surely. Moreover, $X$ is a martingale with respect to the filtration $\{\mathcal{F}_t, t \ge 0\}$.

Constraint (\ref{eqn:approx-cons2.3}) relaxes the requirement that $\delta_k(t) \in \{ 0, 1 \}$ to allow $\delta_k(t) \in [ 0, 1 ]$. As the reader will see below, this relaxation is immaterial to our results because the optimal policy we propose sets $\delta_k(t) \in \{ 0, 1 \}$ for all $k, t$. But, it simplifies the analysis.

\begin{figure}[htbp]
\centering
\includegraphics[scale=0.65]{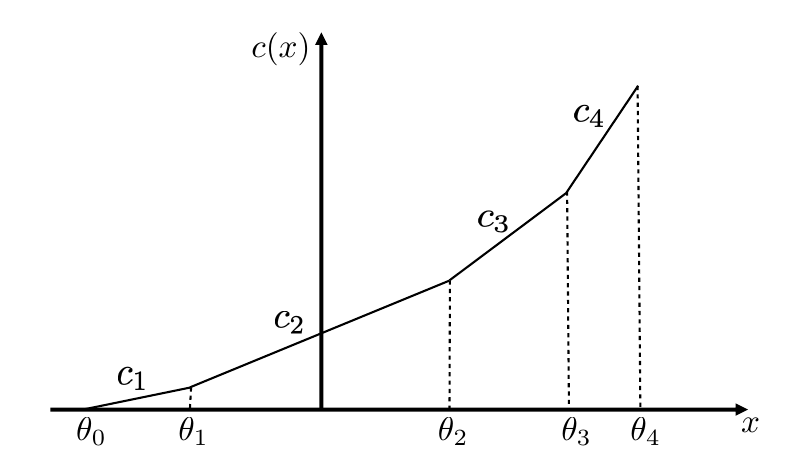}
\caption{An illustrative $c(\cdot)$ function with $K=4$.}
\label{fig:cx}
\end{figure}

To simplify the formulation further, let $A = [ \theta_0, \theta_0 + \sum_{k=1}^K \mu_k]$, and define for $x \in A$,
\begin{align}
c(x) = \min_{\delta_k} \left \{ 
     \sum_{k=1}^K  c_k \mu_k \delta_k : \theta_0 + \sum_{k=1}^K \mu_k \delta_k = x, 
     \,\, 0 \le \delta_k \le 1, \,\, k=1, ... , K
\right\}. \nonumber
\end{align}
Letting $\theta_0 = \theta$, and $\theta_k = \theta + \sum_{l=1}^k \mu_l$ for $k = 1, ... , K$, it is straightforward to show that 
\begin{align}
c(x) = \left\{ \begin{array}{ll}
     c_1(x - \theta_0), & \mbox{if }\,\, \theta_{0} < x \le \theta_1,  \\
     \sum_{i = 1}^{k-1} c_i(\theta_i - \theta_{i-1}) + c_k (x- \theta_{k-1}),
     & \mbox{if }\,\, \theta_{k-1} < x \le \theta_k, \,\, k=2, ... , K.
     \end{array}
\right.  \label{eqn:cx}
\end{align}
Figure~\ref{fig:cx} shows an illustrative $c(\cdot)$ function with $K=4$.

Using this cost function, consider the following formulation: Choose the drift rate process $\theta(\cdot)$ taking values in A so as to 
\begin{align}
& \min \limsup_{t \rightarrow \infty} \frac{1}{t} \mathbb{E} \left[
      \int_{0}^{t} c(\theta(s)) ds
     + h \int_{0}^{t}  Z(s) ds + p L(t)
     \right] \label{eqn:appox-obj3} \\ 
& \mbox{such that} \nonumber \\ 
&  Z(t) = X(t) + \int_{0}^{t} \theta(s) ds + L(t), \,\, t \ge 0, \label{eqn:approx-cons3.1} \\
&  \,\,  Z(t) \ge 0, \,\, t \ge 0, \label{eqn:approx-cons3.2} \\
&  \,\, \int_0^{\infty} 1_{\{Z(s) > 0\}} dL(s) = 0,  \label{eqn:approx-cons3.3} \\
&  \,\, \theta(\cdot), L \ \mbox{ are adapted to } \{ \mathcal{F}_t, t \ge 0 \}. \label{eqn:approx-cons3.4} 
\end{align}

\begin{figure}[htbp]
\centering
\subfigure[An illustrative $\psi(\cdot)$ function with $K=4$.]{\label{fig:psi}\includegraphics[scale=0.6]{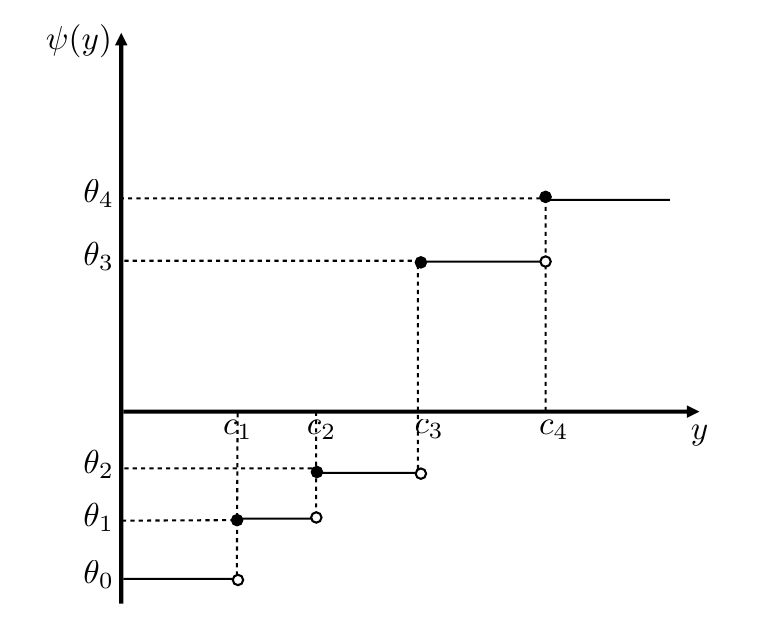}} 
\subfigure[An illustrative $\phi(\cdot)$ function with $K=4$.]{\label{fig:phi}\includegraphics[scale=0.6]{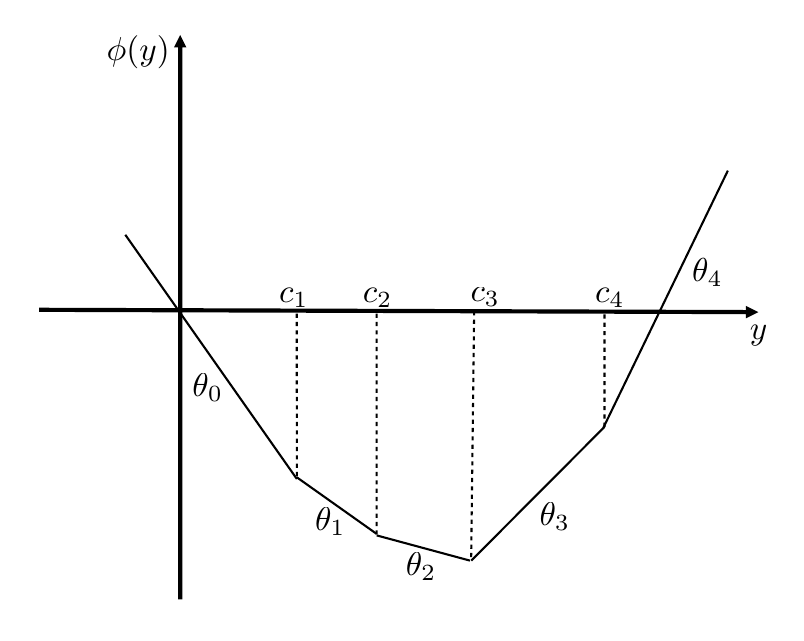}} 
\caption{Illustrative $\psi(\cdot)$ and $\phi(\cdot)$ functions.}
\label{fig:psiphi}
\end{figure}

The following proposition establishes the equivalence of the formulations (\ref{eqn:appox-obj2})-(\ref{eqn:approx-cons2.5}) and (\ref{eqn:appox-obj3})-(\ref{eqn:approx-cons3.4}). 
\begin{proposition} \label{prop:equivform}
Formulations (\ref{eqn:appox-obj2})-(\ref{eqn:approx-cons2.5}) and (\ref{eqn:appox-obj3})-(\ref{eqn:approx-cons3.4}) are equivalent in the following sense: For any feasible policy $\delta(\cdot)$ of formulation (\ref{eqn:appox-obj2})-(\ref{eqn:approx-cons2.5}), there exists a feasible policy $\theta(\cdot)$ of formulation (\ref{eqn:appox-obj3})-(\ref{eqn:approx-cons3.4}) whose average cost is less than or equal to that of $\delta(\cdot)$. Similarly, for any feasible policy $\theta(\cdot)$ of formulation (\ref{eqn:appox-obj3})-(\ref{eqn:approx-cons3.4}), there is a feasible poligy $\delta(\cdot)$ of formulation (\ref{eqn:appox-obj2})-(\ref{eqn:approx-cons2.5}) that has the same average cost. Moreover, the two formulations have the same average cost.
\end{proposition}
Henceforth, we shall focus attention on formulation (\ref{eqn:appox-obj3})-(\ref{eqn:approx-cons3.4}).
To facilitate the analysis, define the convex conjugate of $c(\cdot)$ as follows:
\begin{align}
\phi(y) = \max_{x \in A} \{ yx - c(x) \}, \,\, y \in \mathbb{R}. \label{eqn:phiconcon}
\end{align}
Note that $\phi(\cdot)$ is Lipschitz continuous, as stated in the following lemma.
\begin{lemma} \label{lem:philip}
The function $\phi(\cdot)$ is Lipschitz continuous with Lipschitz constant $L = \max\{ | \theta_k | : k = 0, 1, ... , K\}$.
\end{lemma}
It is straightforward to argue that $\phi(y)$ is well-defined and finite for all $y \in \mathbb{R}$, and that there is a smallest $\psi(y) \in A$ achieving the maximum. That is, the following is well defined, too.
\begin{align*}
 \psi(y) = \inf \arg \max_{x \in A} \{ yx - c(x) \}, \,\,\, y \in \Re. 
\end{align*}
It is straightforward to show that
\begin{align}
\psi(y) = \left\{ \begin{array}{ll}
     \theta_0 & \mbox{if } y \le c_1,  \\
     \theta_{k-1} & \mbox{if } c_{k-1} < y \le c_k, \,\, k=2, ... , K, \\
     \theta_K & \mbox {if } y > c_K, 
     \end{array}
\right.  \label{eqn:psi2} 
\end{align}
and
\begin{align}
\phi(y) = \left\{ \begin{array}{ll}
     \theta_0 y & \mbox{if } y \le c_1, \\
     \theta_{k-1} y - c(\theta_{k-1}) & \mbox{if } c_{k-1} < y \le c_k, \,\, k=2, ... , K, \\
     \theta_K y - c(\theta_K) & \mbox {if } y > c_K.
     \end{array} 
\right.  \label{eqn:phi2} 
\end{align}
Figures \ref{fig:psi} and \ref{fig:phi} display illustrative $\psi(\cdot)$ and $\phi(\cdot)$ functions, respectively. It is immediate from (\ref{eqn:psi2}) and (\ref{eqn:phi2}) that $\phi(y) = \int_0^y \psi(u) du$ for $y \in \mathbb{R}$.

The next section solves the diffusion control problem stated in (\ref{eqn:appox-obj3})-(\ref{eqn:approx-cons3.4}).

%%%%%%%%%%%%%%%%%%%%%%%%%%%%%%%%%%
% SOLUTION TO THE DIFFUSION CONTROL PROBLEM
%%%%%%%%%%%%%%%%%%%%%%%%%%%%%%%%%%
\section{Solution to the Diffusion Control Problem} \label{sec:analysis}
%\input{vol-analysis}

%%%%%% Analysis

In what follows, we restrict attention to stationary Markov policies. Namely, the drift rate chosen at time $t$ depends only on the current queue length $Z(t)$. Therefore, an admissible policy is a function $\theta(\cdot):[0, \infty) \rightarrow A$. As a preliminary to introducing the Bellman equation, we define $C^2[0, \infty)$ as the space of functions $f:[0, \infty) \rightarrow \Re$ that are twice continuously differentiable up to the boundary. Similarly, we define $C^1[0, \infty)$ as the space of functions that are continuously differentiable up to the boundary.

Consider the Bellman equation associated with formulation (\ref{eqn:appox-obj3})-(\ref{eqn:approx-cons3.4}): Find a constant $\beta$ and a function $f \in C^2 [0, \infty)$ that jointly solve the following:
\begin{align}
& \beta = \min_{x \in A} \{ \frac{1}{2} \sigma^2 f''(z) + x f'(z) + c(x) + hz \} \label{eqn:bell1} \\
& \mbox{s.t. } f'(0) = -p \mbox{ and } f' \mbox{ increasing, grows linearly as } x \rightarrow \infty, \ f'(x) \sim \frac{hx}{|\theta_0|} \label{eqn:bell1cons}.
\end{align}
To simplify the Bellman equation, let $g(z) = f(z) + pz$. Substituting this into Equations (\ref{eqn:bell1})-(\ref{eqn:bell1cons}) and rearranging the terms gives the following:
\begin{align}
& \beta = \frac{1}{2} \sigma^2 g''(z) + hz+ \min_{x\in A}
     \{ x(g'(z) - p) + c(x)\} \label{eqn:bell2} \\
& \mbox{s.t. } g'(0) = 0 \mbox{ and } g' \mbox{ increasing, grows linearly  }, g'(x) \sim \frac{hx}{|\theta_0|}. \label{eqn:bell2cons} 
\end{align}
Note that the Bellman equation does not involve the unknown function $g$ itself. So it is really a first-order differential equation. Letting $v(z) = g'(z)$ and using $\phi$, we can rewrite the Bellman equation more succinctly as follows: Find a constant $\beta$ and a function $v \in C^1[0, \infty)$ that jointly solve the following:
\begin{align}
& \beta = \frac{1}{2} \sigma^2 v'(z) + hz - \phi(p - v(z)) \label{eqn:bell3} \\
& \mbox{s.t. } v(0) = 0 \mbox{ and } v \mbox{ increasing, grows linearly}, v(x) \sim \frac{hx}{|\theta_0|}. \label{eqn:bell3cons} 
\end{align}
The solution to the Bellman equation is derived in Subsection \ref{sec:bellman-solution}. Next, we propose a candidate policy given that solution and prove its optimality in Section \ref{sec:cand-policy}. This policy will be characterized further in Section \ref{sec:opt-thresholds}.

As the reader will see below, the function $\psi$ efficiently captures all aspects of the cost function $c(\cdot)$ that are relevant for our purposes. Moreover, the optimal policy will be characterized as follows:
\begin{align}
\theta^*(z) = \psi(p - v(z)), \,\, z \ge 0,  \label{eqn:thetastar}
\end{align}
which can be expressed as a  nested-threshold policy; see Subsection \ref{sec:cand-policy}.

\begin{figure}[htbp]
\centering
\includegraphics[scale=0.5]{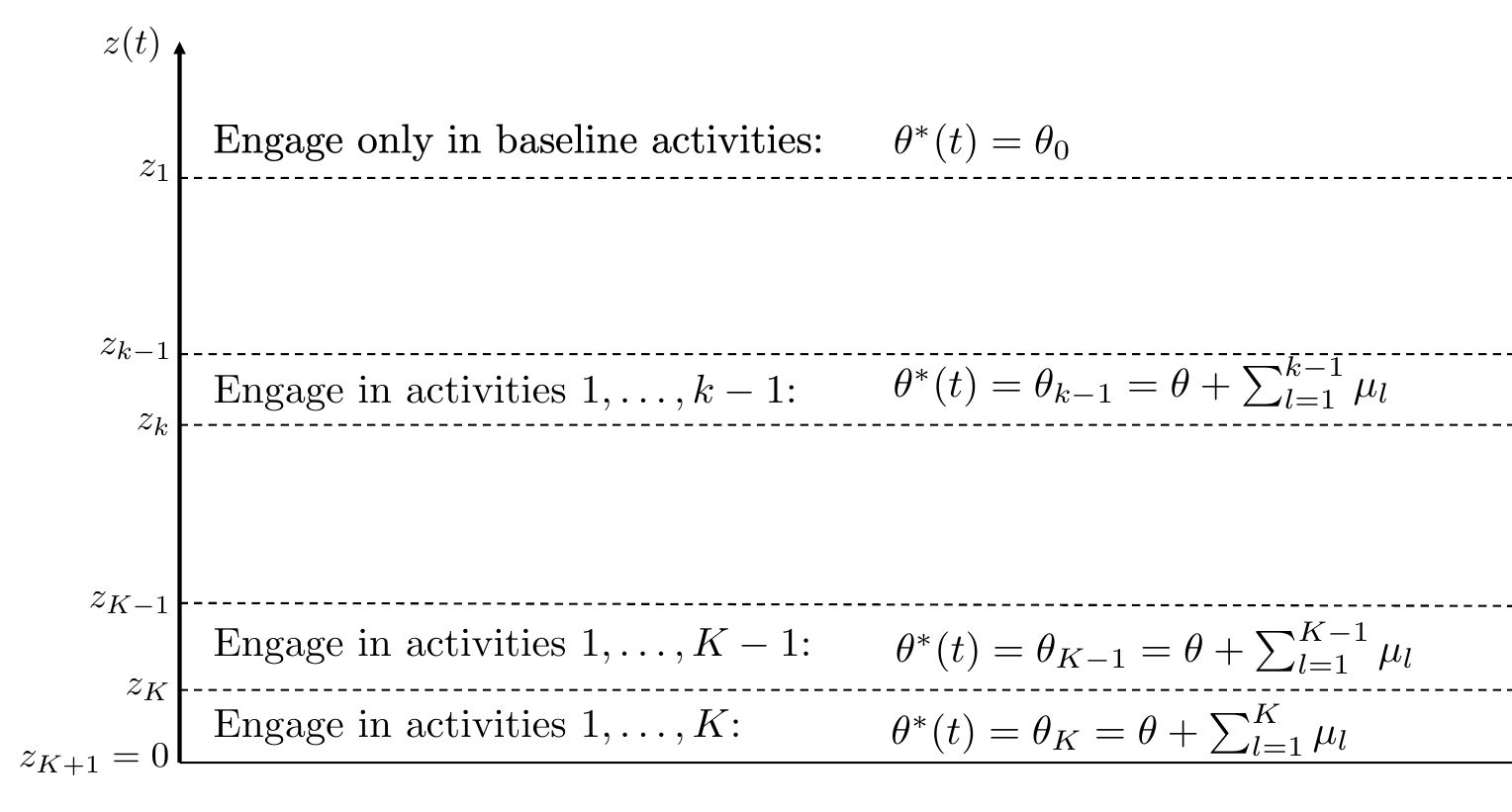}
\caption{The structure of the optimal nested-threshold policy.}
\label{fig:policy}
\end{figure}

As a preliminary to proving that this candidate policy is optimal, next we define the class of admissible policies. This definition simply ensures that the system workload is stable under an admissible policy.

\paragraph{Definition (Admissible Policies).} Let $j^* = \max \{ 0, 1, \dots, K : \theta_j < 0 \}$. Clearly, $\theta_{j^*} < 0$. A policy $\theta(\cdot)$ is admissible if it is stationary Markov and if there exists $\bar{z}$ such that $\theta(z) \le \theta_{j^*} < 0$ for $z \ge \bar{z}$. 

Recall that $\theta_0 < 0$, which ensures there exists an admissible policy.

%%%%%%%%%%%%%%%%%%%%%%%%%%%%%%%%%%
% SOLUTION TO THE BELLMAN EQUATION
%%%%%%%%%%%%%%%%%%%%%%%%%%%%%%%%%%
%3.1
\subsection{Solution to the Bellman Equation} \label{sec:bellman-solution}
%\input{vol-proof-bellman}

%The analysis in this appendix solves the Bellman equation and verifies the optimality of the proposed policy in (\ref{eqn:thetastar}).

To facilitate the analysis to follow, let $\betainf = - \inf \{ \phi(y) \} \ge 0$ and for $\beta > \betainf$, consider the following initial value problem, denoted by IVP($\beta$), that is closely related to the Bellman equation:
%
\begin{align}
\beta &= \frac{1}{2} \sigma^2 v'(x) + hx - \phi(p - v(x)), \,\, x\ge 0, \label{eqn:ivp1} \\
v(0) &= 0. \label{eqn:ivp2}
\end{align}

The following lemma establishes the existence and uniqueness of the solution to the IVP($\beta$). Its proof is standard, and thus, omitted; see for example Theorem $1.1.1$ of \citet{Keller_2018} for a version of this result on a bounded interval, which can easily be extended to the positive real line.

%\label{eqn:ivp} \tag{IVP$(\beta$)} 
%%%%%%%%%%%%%%%%
% LEMMA
%%%%%%%%%%%%%%%%
\begin{lemma}
The initial value problem, IVP($\beta$), stated in (\ref{eqn:ivp1})-(\ref{eqn:ivp2}) has a unique continuously differentiable solution, denoted by $v_{\beta}(x)$, for $\beta > \betainf$.
\end{lemma}
%
%\begin{proof}
%Because $\phi$ is Lipschitz continuous (see Lemma~\ref{lem:philip}), it follows by Picard's iteration arguments (see pages 89-98 of {\color{red} ***** CITE Boyce and DiPrima (1992) *****}) that there exists a $\delta > 0$ such that we have a unique continuously differentiable solution $v_{\beta}$ on $[0, \delta]$. This result can be extended to the entire interval $[0, k]$ for all $k>0$ (and hence to $[0, \infty)$) by mimicking the arguments on page~192 of {\color{red} ***** CITE Mandl (1968) *****}.
%\end{proof}
In what follows, we denote the solution to the IVP($\beta$) by $v_{\beta}(\cdot)$ and study its properties as $\beta$ varies. Ultimately, we show that there exists a unique $\betastar > \betainf$ such that $v_{\betastar}$ satisfies the growth condition in Equation (\ref{eqn:bell3cons}) and thus, ($\betastar, v_{\betastar}$) solve the Bellman equation. Indeed, much of the technical subtlety of the analysis stems from verifying this growth condition, which is crucially used in the proof of optimality of the candidate policy given in Equation (\ref{eqn:thetastar}).

Lemmas~\ref{lem:vbeta} and \ref{lem:beta} are auxiliary results for the analysis to follow. Lemma~\ref{lem:gronwall} is the Gronwall's Inequality, which is used in establishing the continuity part of Lemma~\ref{lem:vbeta}. It is stated for completeness; for a proof, see, for example, \citet{ata-etal-2019}. Lemma~\ref{lem:vbeta} helps us study how the solution $v_{\beta}(\cdot)$ of IVP($\beta$) varies with $\beta$ and is proved in Appendix \ref{app:proofs}.

%%%%%%%%%%%%%%%%
% LEMMA lem:gronwall
%%%%%%%%%%%%%%%%
\begin{lemma} \label{lem:gronwall}
Let $F$ be a non-negative function such that
\begin{align}
F(x) \le C + A \int_{a}^{x} F(y) dy, \,\,\, \mbox{for } a \le x \le b \nonumber 
\end{align}
for some constants $A, C > 0$. Then $F(x) \le C e ^{A(x-a)} \,\,\, \mbox{for } x \in [a, b]$. 
\end{lemma}

%%%%%%%%%%%%%%%%
% LEMMA lem:vbeta
%%%%%%%%%%%%%%%%

\begin{lemma} \label{lem:vbeta}
For $\beta > \betainf$, $v_{\beta}(x)$ is increasing and continuous in $\beta$.
\end{lemma}

To facilitate the analysis, define the following two subsets of $(\betainf, \infty)$:
\begin{align}
\mathcal{I} &= \{ \beta > \betainf : v_{\beta} \mbox{ increases strictly to } \infty \} \nonumber\\
\mathcal{N} &= \{ \beta > \betainf : v_{\beta} \mbox{ is first increasing, then it is decreasing } \} \nonumber
\end{align}

%%%%%%%%%%%%%%%%
% LEMMA
%%%%%%%%%%%%%%%%

\begin{lemma} \label{lem:beta}
For each $\beta > \betainf$, we have either $\beta \in \mathcal{I}$ or $\beta \in \mathcal{N}$, i.e. $\mathcal{I} \cup \mathcal{N} = (\betainf, \infty)$ and $\mathcal{I} \cap \mathcal{N} = \emptyset$. Moreover, for $\beta_2 > \beta_1 > \betainf$, if $\beta_1 \in \mathcal{I}$, then $\beta_2 \in \mathcal{I}$.
\end{lemma}
%

%%%%%%%%%%%%%%%%
% LEMMA
%%%%%%%%%%%%%%%%
\begin{lemma} \label{lem:INnotempty}
The sets $\mathcal{I}$, $\mathcal{N}$ are nonempty. That is, $\mathcal{I} \not= \emptyset$, $\mathcal{N} \not= \emptyset$. Moreover, $\left( p | \theta_0 | + \frac{h \sigma^2 }{2 | \theta_0| }, \infty \right) \subset \mathcal{I}$.
\end{lemma}

Letting $\betastar = \inf\mathcal{I}$, the next lemma shows $\betastar \in \mathcal{I}$.

%%%%%%%%%%%%%%%%
% LEMMA
%%%%%%%%%%%%%%%%
\begin{lemma} \label{lem:betastar}
We have that $\betastar \in \mathcal{I}$.
\end{lemma}

\begin{proof}
Suppose not. Then $v_{\betastar}$ is first increasing and then decreasing. Let $x_0 = \arg \max v_{\betastar}(x)$ and fix $x_1> x_0$ such that $v_{\betastar}(x_1) < v_{\betastar}(x_0)$, and let $\varepsilon = (v_{\betastar}(x_0) - v_{\betastar}(x_1))/2 >0$. Because $v_{\betastar}(x_1)$ is continuous and increasing in $\beta$, there exists $\beta' > \betastar$ such that $\beta' \in \mathcal{I}$ and
\begin{align}
v_{\betastar}(x_1) \le v_{\beta'}(x_1) \le v_{\betastar}(x_1) + \varepsilon \le v_{\betastar}(x_0) - \varepsilon. 
    \label{eqn:lembetastar1}
\end{align}
Once again, because $v_{\beta}(x)$ is increasing in $\beta$, we also have
\begin{align}
v_{\beta'}(x_0) \ge v_{\betastar}(x_0). \label{eqn:lembetastar2}
\end{align}
Combining (\ref{eqn:lembetastar1}) and (\ref{eqn:lembetastar2}) gives
$v_{\beta'}(x_1) \le v_{\beta'}(x_0) - \varepsilon$, 
which contradicts that $\beta' \in \mathcal{I}$, in particular, that $v_{\beta'}(\cdot)$ is increasing.
\end{proof}
%%%%%%%%%%%%%%%%%%%%
% COROLLARY
%%%%%%%%%%%%%%%%%%%%
The following is a corollary of Lemmas~\ref{lem:beta}, \ref{lem:INnotempty}, and~\ref{lem:betastar}.
\begin{corollary} \label{cor:betastar}
We have that $\betastar \in \Big( \betainf, p |\theta_0| + \frac{h \sigma^2}{2 |\theta_0|} \Big]$.
\end{corollary}

The next result establishes the desired growth condition, paving the way to solving the Bellman equation.

%%%%%%%%%%%%%%%%
% LEMMA
%%%%%%%%%%%%%%%%
\begin{lemma} \label{lem:betastarlin}
The function $v_{\betastar}(\cdot)$ grows linearly. In particular, $v_{\betastar}(x) \sim \frac{h}{| \theta_0 |}x \quad \mbox{as } x \rightarrow \infty$.
\end{lemma}

\begin{proof}
First note that because $v_{\beta}(x) \rightarrow \infty$ as $x \rightarrow \infty$ for $\beta \in \mathcal{I}$, there exists $x_0(\beta) > 0$ such that $v_{\beta}(x) \ge 2p$ for $x \ge x_0$. Thus,
\begin{align*}
\phi(p - v_\beta(x)) = \theta_0 p -\theta_0 v_\beta(x), \,\,\, x \ge x_0(\beta).
\end{align*}
Fix such an $x_0$ and note that Equation~(\ref{eqn:ivp1}) simplifies to the following:
\begin{align*}
\beta = \frac{\sigma^2}{2} v_{\beta}'(x) + hx -\theta_0 p + \theta_0 v_{\beta}(x).
\end{align*}
Rearranging terms, we write
\begin{align*}
v_{\beta}'(x) + \frac{2 \theta_0 }{\sigma^2}v_{\beta}(x) = 
     \frac{2(\beta + \theta_0 p)}{\sigma^2} - \frac{2hx}{\sigma^2} .
\end{align*}
Multiplying both sides by $\exp\{ \frac{2 \theta_0 x}{\sigma^2} \}$ and integrating over $[x_0, x]$ give the following:
\begin{align*}
\left[ \exp \left\{ \frac{2 \theta_0 x}{\sigma^2} \right\} v(x) \right]' 
     =& \exp \left\{ \frac{2 \theta_0 x}{\sigma^2} \right\} \frac{2(\beta + \theta_0 p)}{\sigma^2} 
     - \exp \left\{ \frac{2 \theta_0 x}{\sigma^2} \right\} \frac{2hx}{\sigma^2} \\
\exp \left\{ \frac{2 \theta_0 x}{\sigma^2} \right\} v(x) - \exp \left\{ \frac{2 \theta_0 x}{\sigma^2} \right\}  v(x_0)
     =& \frac{2(\beta + \theta_0 p)}{\sigma^2} \frac{\sigma^2}{2 \theta_0} \exp \left\{ \frac{2 \theta_0 x}{\sigma^2} \right\} \Bigg|_{x_0}^x \\
     &- \frac{2h}{\sigma^2} \int_{x_0}^{x} x \exp \left\{ \frac{2 \theta_0 x}{\sigma^2} \right\} dx.
\end{align*}
That is, the following holds:
\begin{align}
&v(x) \exp \left\{ \frac{2 \theta_0 x}{\sigma^2} \right\}  
     - v(x_0) \exp \left\{ \frac{2 \theta_0 x_0}{\sigma^2} \right\} \nonumber \\
&\quad = \left( \frac{p}{\theta_0} + p \right) \left[ \exp \left\{ \frac{2 \theta_0 x}{\sigma^2} \right\}
     - \exp \left\{ \frac{2 \theta_0 x_0}{\sigma^2} \right\} \right] - I, \label{eqn:lembetastarlin1}
\end{align}
where $I = \frac{2h}{\sigma^2} \int_{x_0}^{x} z  \exp\left\{ \frac{2 \theta_0 z}{\sigma^2} \right\} dz$. We calculate $I$ by integration by parts:
\begin{align*}
I &= \frac{2h}{\sigma^2} \frac{\sigma^2}{2 \theta_0} z \exp \left\{ \frac{2 \theta_0 z }{\sigma^2} \right\} \Bigg|_{x_0}^{x}
     - \frac{2h}{\sigma^2} \int_{x_0}^x \frac{\sigma^2}{2 \theta_0}
     \exp \left\{ \frac{2 \theta_0 z }{\sigma^2} \right\} dz \\
&= \frac{h}{\sigma^2} z \exp \left\{ \frac{2 \theta_0 z }{\sigma^2} \right\} \Bigg|_{x_0}^{x}
     - \frac{h}{\theta_0}  \int_{x_0}^x \exp \left\{ \frac{2 \theta_0 z }{\sigma^2} \right\}  dz \\
&= \frac{h}{\theta_0} x \exp \left\{ \frac{2 \theta_0 x }{\sigma^2} \right\}
     - \frac{h}{\theta_0} x_0 \exp \left\{ \frac{2 \theta_0 x_0 }{\sigma^2} \right\}
     - \frac{h}{\theta_0} \frac{\sigma^2}{2 \theta_0 } \exp \left\{ \frac{2 \theta_0 z}{\sigma^2} \right\} \Bigg|_{x_0}^{x} \\
&= \frac{h}{\theta_0} x \exp \left\{ \frac{2 \theta_0 x}{\sigma^2} \right\}
     - \frac{h}{\theta_0} x_0 \exp \left\{ \frac{2 \theta_0 x_0 }{\sigma^2} \right\}
     - \frac{h \sigma^2}{2 \theta_0^2} \exp \left\{ \frac{2 \theta_0 z }{\sigma^2} \right\}
     + \frac{h \sigma^2}{2 \theta_0^2} \exp \left\{ \frac{2 \theta_0 x_0 }{\sigma^2} \right\}.
\end{align*}
Substituting this into Equation~(\ref{eqn:lembetastarlin1}) gives 
\begin{align*}
v_{\beta}(x) \exp \left\{ \frac{2 \theta_0 x }{\sigma^2} \right\} 
     =& v_{\beta}(x_0) \exp \left\{ \frac{2 \theta_0 x_0 }{\sigma^2} \right\} 
     + \left( \frac{p}{\theta_0} +p \right) 
     \left[ \exp \left\{ \frac{2 \theta_0 x }{\sigma^2} \right\} - \exp \left\{ \frac{2 \theta_0 x_0 }{\sigma^2} \right\}\right] \\
& -\frac{hx }{ \theta_0}  \exp \left\{ \frac{2 \theta_0 x }{\sigma^2} \right\} 
     + \frac{h x_0}{ \theta_0}  \exp \left\{ \frac{2 \theta_0 x_0 }{\sigma^2} \right\} 
     + \frac{h \sigma^2}{2 \theta_0^2} \exp \left\{ \frac{2 \theta_0 x }{\sigma^2} \right\} 
     - \frac{h \sigma^2}{2 \theta_0^2} \exp \left\{ \frac{2 \theta_0 x_0 }{\sigma^2} \right\}.
\end{align*}
Rearranging the terms gives 
\begin{align*}
v_{\beta}(x) \exp \left\{ \frac{2 \theta_0 x }{\sigma^2} \right\} 
     =& \exp \left\{ \frac{2 \theta_0 x_0 }{\sigma^2} \right\} \left[ v_{\beta}(x_0)- \left( \frac{p}{\theta_0} +p \right)
     + \frac{h x_0}{ \theta_0} - \frac{h \sigma^2}{2 \theta_0^2} \right] \\
 &+ \exp \left\{ \frac{2 \theta_0 x }{\sigma^2} \right\} \left[ \frac{\beta}{\theta_0} + p
      -\frac{hx}{\theta_0} + \frac{h \sigma^2}{2 \theta_0^2} \right].
\end{align*}
That is, we have that
\begin{align}
v_{\beta}(x) = \exp\left\{ -\frac{2 \theta_0( x - x_0)}{\sigma^2} \right\}
     \left( v_{\beta}(x_0) - \left( \frac{\beta}{\theta_0} + p \right) + \frac{h x_0}{\theta_0}
     - \frac{h \sigma^2}{2 \theta^2} \right) \nonumber \\
     + \left( \frac{\beta}{\theta_0} + p \right)
     +  \frac{h x_0}{|\theta_0|} + \frac{h \sigma^2}{2 \theta_0^2}. \label{eqn:lembetastarlin2}
\end{align}

We claim that $x_0(\betastar)$ is such that
\begin{align}
v_{\betastar}(x_0(\betastar)) - \left( \frac{\betastar}{\theta_0} + p \right)
     + \frac{h }{\theta_0} x_0(\betastar) - \frac{h \sigma^2}{2 \theta_0^2} = 0, \label{eqn:lembetastarlin3}
\end{align}
so that
\begin{align}
v_{\betastar}(x) = \left( \frac{\beta}{\theta_0} + p \right) +  \frac{h x}{|\theta_0|}
      +\frac{h \sigma^2}{2 \theta_0^2} \quad \mbox{for } x > x_0(\betastar),  \label{eqn:lembetastarlin4}
\end{align}
which proves the statement. Suppose that (\ref{eqn:lembetastarlin3}) does not hold. Then we must have
\begin{align}
v_{\betastar}(x_0(\betastar)) - \left( \frac{\betastar}{\theta_0} + p \right)
     + \frac{h }{\theta_0} x_0(\betastar) - \frac{h \sigma^2}{2 \theta_0^2} > 0. \label{eqn:lembetastarlin5}
\end{align}
Otherwise, i.e., if the left-hand side of Equation~(\ref{eqn:lembetastarlin5}) is negative, then $v_{\betastar(x)} \rightarrow - \infty$ as $ x \rightarrow \infty$, which contradicts that $\betastar \in \mathcal{I}$. However, Equation~(\ref{eqn:lembetastarlin5}) implies by continuity that there exists $\varepsilon > 0$ sufficiently small such that (\ref{eqn:lembetastarlin5}) holds for $\tilde{\beta} = \betastar - \varepsilon$, which, in turn, implies that
\begin{align*}
v_{\beta}(x) \rightarrow \infty \quad \mbox{as } x \rightarrow \infty,
\end{align*}
i.e., that $\tilde{\beta} \notin \mathcal{N}$. Thus, we must have $\tilde{\beta} = \betastar - \varepsilon \in \mathcal{I}$, which contradicts that $\betastar = \inf \mathcal{I}$. Therefore, (\ref{eqn:lembetastarlin3}) - (\ref{eqn:lembetastarlin4}) holds.
\end{proof} \vspace{-20pt}

The following corollary is then immediate.

%%%%%%%%%%%%%%%%%%
% COROLLARY
%%%%%%%%%%%%%%%%%%
\begin{corollary} \label{cor:betastar-pair}
The pair ($\betastar, v_{\betastar}$) solve (\ref{eqn:bell3})-(\ref{eqn:bell3cons}).
\end{corollary}

Lastly, to provide a solution of the Bellman equation, define
\begin{align}
f(z) = \int_0^z v_{\betastar}(s) ds - pz, \,\,\, z \ge 0. \label{eqn:propbellsoln1}
\end{align}

%%%%%%%%%%%%%%%%%%%%
% Proposition
%%%%%%%%%%%%%%%%%%%%
\begin{proposition} \label{prop:bellsoln}
The pair $(\betastar, f)$ solve the Bellman equation (\ref{eqn:bell1})- (\ref{eqn:bell1cons}).
\end{proposition}

%%%%%%%%%%%%%%%%%%%%%%%%%%%%%%%%%%
% THE CANDIDATE POLICY AND ITS OPTIMALITY
%%%%%%%%%%%%%%%%%%%%%%%%%%%%%%%%%%
%3.2
\subsection{The candidate policy and its optimality} \label{sec:cand-policy}

This subsection first establishes the optimality of the candidate policy introduced above. Recall that the proposed policy $\theta^*(\cdot)$ is given as follows:
\begin{align}
\theta^*(z) = \psi(p - v_{\betastar}(z)), \,\,\, z \ge 0. \label{eqn:propbellsoln2}
\end{align}
Theorem~\ref{thm:optpol} below verifies the optimality of this candidate policy. As a preliminary to its proof, the following lemma proves useful properties of the function $f$ defined in Equation (\ref{eqn:propbellsoln1}); see Appendix \ref{app:proofs} for its proof.

%%%%%%%%%%%%%%%%%%%%%%%
% LEMMA
%%%%%%%%%%%%%%%%%%%%%%%
\begin{lemma} \label{lem:propertiesf}
Under any admissible policy, we have that
\begin{align*}
&\mbox{i) } \mathbb{E} \left[  \int_0^t f'(Z(t)) dB(t) \right] = 0, \,\,\, t\ge0 \\
&\mbox{ii) } \lim_{t \rightarrow \infty} \frac {\mathbb{E} \left[ f(Z(t)) \right]}{t} = 0.
\end{align*}
\end{lemma}

%%%%%%%%%%%%%%%%%%
% THEOREM
%%%%%%%%%%%%%%%%%%
\begin{theorem} \label{thm:optpol}
The candidate policy $\theta^*(z) = \psi(p - v_{\betastar}(z))$ for $z \ge 0$ is optimal, and its long-run average cost is $\betastar$.
\end{theorem}

\begin{proof}
Note from Equations~(\ref{eqn:bell1})-(\ref{eqn:bell1cons}) and~(\ref{eqn:thetastar}) that the candidate policy satisfies the following:
\begin{align}
\betastar = \frac{1}{2} \sigma^2 f''(z) + hz + \theta^*(z) f'(z) + c(\theta^*(z)). \label{eqn:thmoptpol1}
\end{align}
Similarly, for an arbitrary admissible policy $\theta(\cdot)$, we have
\begin{align}
\betastar \le \frac{1}{2} \sigma^2 f''(z) + hz + \theta(z) f'(z) + c(\theta(z)). \label{eqn:thmoptpol2}
\end{align}
Also, for any admissible policy $\theta(\cdot)$, applying Ito's lemma to $f(Z(t))$ gives
\begin{align}
f(Z(t)) - f(Z(0)) =& \int_0^t [ \theta(z(t)) f'(Z(t)) + \frac{\sigma^2}{2} f''(Z(t))] dt \nonumber\\
&+ \int_0^t \sigma f'(Z(t)) dB(t) + f'(0) L(t), \label{eqn:thmoptpol3}
\end{align}
see Chapters~4 and~6 of \citet{harrison-2013}. By taking the expectations of both sides of Equation~(\ref{eqn:thmoptpol3}), it follows from Lemma~\ref{lem:propertiesf} that 
\begin{align}
\mathbb{E} [f(Z(t)) ] - \mathbb{E} [f(Z(0)) ]  
     = \mathbb{E}  \int_0^t ( \theta(z(t)) f'(Z(s)) + \frac{\sigma^2}{2} f''(Z(s)) ds 
     + f'(0) \mathbb{E} [ L(t) ]. \label{eqn:thmoptpol4}
\end{align}

For any admissible policy $\theta(\cdot)$, combining Equations~(\ref{eqn:thmoptpol2}) and~(\ref{eqn:thmoptpol4}) gives
\begin{align*}
\mathbb{E} f(Z(t)) - \mathbb{E} f(Z(0)) \ge \mathbb{E}  \left[ \int_0^t (\betastar - c(\theta(Z(s)) - hZ(s)) ds \right]
     -p \mathbb{E} [ L(t)].
\end{align*}
Rearranging the terms gives
\begin{align*}
\mathbb{E}  \left[ \int_0^t  c(\theta(Z(s))) ds + \int_0^t  h(Z(s)) ds + p L(t)  \right]
     \ge \betastar t +\mathbb{E} f(Z(0)) - \mathbb{E} f(Z(t)).
\end{align*}
Combining this with Lemma~\ref{lem:propertiesf} gives
\begin{align*}
\underline{\lim}_{t \rightarrow \infty} \frac{1}{t} \mathbb{E} \left[ \int_0^t c(\theta(Z(s))) 
     +h Z(s) ] ds + p L(t) \right] \ge \betastar.
\end{align*}
Similarly, for candidate policy $\theta^*(\cdot)$, combining Equations~(\ref{eqn:thmoptpol1}) and~(\ref{eqn:thmoptpol4}) gives
\begin{align*}
\mathbb{E} \left[ \int_0^t c(\theta(Z(s))) ds + \int_0^t  h Z(s) ds + p L(t) \right]
     = \betastar t + \mathbb{E} f (Z(0)) - \mathbb{E} f(Z(t)).
\end{align*}
Then it follows from Lemma~\ref{lem:propertiesf} that
\begin{align*}
\lim_{t \rightarrow \infty} \frac{1}{t} \mathbb{E} \left[ \int_0^t c(\theta(Z(s))) ds
     + \int_0^t h Z(s) ds + p L(t) \right] = \beta^*.
\end{align*}
Therefore, the candidate policy is optimal and its long-run average cost is $\betastar$.
\end{proof} \vspace{-30pt}

%%%%%%%%%%%%%%%%%%%%%%%%%%%%%%%%%%
% FURTHER CHARACTERIZATION OF THE OPTIMAL POLICY
%%%%%%%%%%%%%%%%%%%%%%%%%%%%%%%%%%
%3.3
\subsection{Further Characterization of the Optimal Policy} \label{sec:opt-thresholds}

This subsection characterizes the optimal policy as a nested threshold policy and the corresponding thresholds. Recall from Corollary \ref{cor:betastar-pair} that $v_{\betastar}(\cdot)$ is strictly increasing, and hence, invertible. Denoting that inverse by $v_{\betastar}^{-1}(\cdot)$ and letting $c_{K+1} = p$ for notational convenience, we define the thresholds:
\begin{align}
z_k = v_{\betastar}^{-1}(p-c_k), \ \ \  \,\,\, k = 1, \ldots,  K+1.  \label{eqn:zk}
\end{align}
It follows from the monotonicity of $v_{\betastar}(\cdot)$ that 
$0 = z_{K+1} < z_{K} < \ldots < z_2 < z_1$.
Defining $z_0 = \infty$ for notational convenience, it follows from Equation (\ref{eqn:psi2}) and (\ref{eqn:propbellsoln2}) that the candidate policy $\theta^*(\cdot)$ satisfies the following:
\begin{align}
\theta^*(z) = \theta_{k-1} \ \ \mbox{for}  \,\, z_{k} \leq z < z_{k-1}, \  k = 1, \ldots,  K+1. \label{eqn:thetastarcandidate}
\end{align}

In other words, the candidate policy is a nested-threshold policy, see Figure \ref{fig:policy}. The following corollary is immediate from Theorem~\ref{thm:optpol}.

\begin{corollary} \label{cor:threshold-policy}
The nested-threshold policy is given in Equation (\ref{eqn:zk}) is optimal.
\end{corollary}

In order to facilitate the computation of the thresholds $z_1, \ldots, z_K$, we next provide a closed-form formula for $v_{\betastar}(\cdot)$; and the characterization of $z_1, \ldots, z_K$ amounts to inverting this function, cf. Equation (\ref{eqn:zk}). For notational brevity, we define $u_k(\cdot)$ for $k=1,\ldots, K+1$ as follows:
\begin{align}
u_k(x) =&  \exp \left\{ - \frac{2 \theta_{k-1}}{\sigma^2} (x - z_k) \right\} 
     \left( p - c_k - \frac{\beta + p\theta_{k-1} - c(\theta_{k-1}) }{\theta_{k-1}} + \frac{h z_k}{\theta_{k-1}}
     - \frac{h \sigma^2}{2 \theta_{k-1}} \right) \nonumber \\
&+  \left(\frac{\beta + p\theta_{k-1} - c(\theta_{k-1}) }{\theta_{k-1}} - \frac{hx}{\theta_{k-1}} +\frac{h \sigma^2}{ 2 \theta_{k-1}^2} \right), \ x\geq 0, \nonumber
\end{align}
which implicitly assumes $\theta_{k-1} \neq 0$. The following proposition characterizes $v_{\betastar}(\cdot)$.

\begin{proposition} \label{prop:valuefunc}
We have the following: 
\begin{align}
v_{\betastar}(x) = \left\{ \begin{array}{ll}
     u_{K+1}(x) & \ \ \ \mbox{if }\, \theta_{K} \neq 0,  \\
     \frac{2}{\sigma^2} (\beta^* - c(\theta_K))x - \frac{h x^2}{\sigma^2} & \ \ \ \mbox{if }\, \theta_{K} = 0
     \end{array}
\right. \label{eqn:vbetastar}
\end{align}
for $x \in [0, z_K]$, where $z_K = v_{\betastar}^{-1}(p-c_K)$. Moreover, for $x \in [z_k, z_{k-1})$ and $k=K, \ldots, 1$, we have that 
\begin{align}
v_{\betastar}(x) = \left\{ \begin{array}{ll}
     u_{k}(x) & \ \ \ \mbox{if }\, \theta_{k-1} \neq 0,  \\
     p-c_k + \frac{2}{\sigma^2} (\beta^* - c(\theta_{k-1}))(x-z_k) - \frac{h (x^2-z_k^2)}{\sigma^2} & \ \ \ \mbox{if }\, \theta_{k-1} = 0,
     \end{array}
\right.  \label{eqn:vbetastar2}
\end{align}
where $z_k = v_{\betastar}^{-1}(p-c_k)$.

\end{proposition} 

\begin{proof}
First, we prove (\ref{eqn:vbetastar}). For $x \in [0, z_K]$, using (\ref{eqn:phi2}), the IVP($\beta$) in Equation (\ref{eqn:ivp1}) can be written more explicitly as follows:  
\begin{align}
\beta = \frac{1}{2} \sigma^2 v'(x) + hx - [\theta_K (p-v(x)) - c(\theta_K)] \label{eqn:ivp-proof1}.
\end{align}

If $\theta_{K} = 0$, then this simplifies to the following:
\begin{align*}
v'(x) =& \frac{2}{\sigma^2} [\beta - c(0)] - \frac{2h}{\sigma^2} x.
\end{align*}

Integrating both sides of this and using the boundary condition $v(0)=0$ yields the result:
\begin{align*}
v(x)=& \frac{2}{\sigma^2} [\beta - c(0)] x - \frac{hx^2}{\sigma^2}.
\end{align*}

Now, consider the case $\theta_{K} \neq 0$. Recall $c_{K+1} = p$ and $z_{K+1} = 0$, and note that $u_{K+1}(x)$ is given as follows:
\begin{align*}
u_{K+1}(x) =  \frac{\beta + p\theta_k - c(\theta_K)}{\theta_K}  - \frac{hx}{\theta_K} + \frac{h \sigma^2}{2 \theta_K^2}
			- \left[ \frac{h \sigma^2}{2 \theta_K^2}  + \frac{\beta + p\theta_k - c(\theta_K)}{\theta_K} \right]
     \exp \left\{- \frac{2 \theta_K x}{\sigma^2} \right\}. \nonumber
\end{align*}

Then, it suffices to show that $v(x) = u_{K+1}(x)$. To this end, rearranging the terms in Equation (\ref{eqn:ivp-proof1}) yields the following:
\begin{align*}
v'(x) + \frac{2 \theta_K}{\sigma^2} v(x) = \frac{2}{\sigma^2} [\beta + p\theta_K - c(\theta_K)] - \frac{2h}{\sigma^2} x.
\end{align*}

Multiplying both sides of the equation with the integrating factor $\exp \left\{ 2 \theta_K x / \sigma^2 \right\}$ gives the following:
\begin{align*}
\left[ \exp \left\{ \frac{2 \theta_K}{\sigma^2} x\right\} v(x) \right]'
     =& \frac{2}{\sigma^2} [ \beta + p\theta_K - c(\theta_K)] \exp \left\{ \frac{2 \theta_K}{\sigma^2} x\right\} 
		- \exp \left\{ \frac{2 \theta_K}{\sigma^2} x\right\} x \frac{2h}{\sigma^2} 
\end{align*}

Then, integrating both sides and using the boundary condition $v(0)$ gives that
\begin{align*}
\exp \left\{ \frac{2 \theta_K}{\sigma^2} x\right\} v(x) 		=&  \frac{\beta + p\theta_k - c(\theta_K)}{\theta_K}  \left[ \exp \left\{ \frac{2 \theta_K}{\sigma^2} x\right\} -1 \right] \\
&- \frac{2h}{\sigma^2} \left[ \frac{\sigma^2}{2 \theta_K} z \exp \left\{ \frac{2 \theta_K}{\sigma^2} z\right\}
     \Bigg|_0^x - \frac{\sigma^2}{2 \theta_K} \int_0^x \exp \left\{ \frac{2 \theta_K}{\sigma^2} z\right\} dz \right] .
\end{align*}

Rearranging the terms on the right hand side and dividing both sides of the equation with $\exp \left\{ 2 \theta_K x / \sigma^2 \right\}$ gives the desired result:	
		\begin{align*}
v(x) =  \frac{\beta + p\theta_k - c(\theta_K)}{\theta_K}  - \frac{hx}{\theta_K} + \frac{h \sigma^2}{2 \theta_K^2}
			- \left[ \frac{h \sigma^2}{2 \theta_K^2}  + \frac{\beta + p\theta_k - c(\theta_K)}{\theta_K} \right]
     \exp \left\{- \frac{2 \theta_K x}{\sigma^2} \right\}. \nonumber
\end{align*}

Next, consider $x \in [z_k, z_{k-1})$ for $k =K,\ldots,1$. 
Then, as above, using (\ref{eqn:phi2}), the IVP($\beta$) given in Equation (\ref{eqn:ivp1}) can be written more explicitly as follows: The function $v(\cdot)$ solves the following
\begin{align}
&\beta = \frac{1}{2} \sigma^2 v'(x) + hx - (\theta_{k-1} (p-v(x)) - c(\theta_{k-1})) \label{eqn:ivp-proof2} \\
&\mbox{s.t.} \,\,\,v(z_k) = p - c_k.   \nonumber
\end{align}

If $\theta_{k-1} = 0$, then this simplifies to the following:
\begin{align*}
v'(x) =& \frac{2}{\sigma^2} [\beta - c(\theta_{k-1})] - \frac{2h}{\sigma^2} x.
\end{align*}

Integrating both sides of this over $[z_k, x]$ and using the boundary condition that $v(z_k) = p- c_k$ gives the desired result.
\begin{align*}
v(x)  =& p - c_{k} + \frac{2}{\sigma^2} [\beta - c(\theta_{k-1})] (x- z_k) - \frac{h}{\sigma^2} (x^2 - z_k^2).
\end{align*}

Now, consider the case $\theta_{k-1} \neq 0$. It suffices to show that $v(x) = u_k(x)$. Proceeding as above and using Equation (\ref{eqn:ivp-proof2}) yields:
\begin{align*}
\exp \left\{ \frac{2 \theta_{k-1}}{\sigma^2} z\right\} v(z) \Bigg|_{z_k}^x
     =& \frac{\beta + p\theta_{k-1} - c(\theta_{k-1}) }{\theta_{k-1}} 
     \exp \left\{ \frac{2 \theta_{k-1}}{\sigma^2} z\right\} \Bigg|_{z_k}^x - \frac{2h}{\sigma^2} \int_{z_k}^x z \exp \left\{ \frac{2 \theta_{k-1}}{\sigma^2} z\right\}  dz.
\end{align*}

Rearranging the terms and using the boundary condition $v(z_k) = p-c_k$ give
\begin{align*}
\exp \left\{ \frac{2 \theta_{k-1}}{\sigma^2} x\right\} v(x) - (p-c_k) \exp \left\{ \frac{2 \theta_{k-1}}{\sigma^2} z_k\right\}
     =& \frac{\beta + p\theta_{k-1} - c(\theta_{k-1}) }{\theta_{k-1}} \left( \exp \left\{ \frac{2 \theta_{k-1}}{\sigma^2} x \right\} - \exp \left\{ \frac{2 \theta_{k-1}}{\sigma^2} z_k \right\} \right) \\
&- \frac{hx}{\theta_{k-1}}  \exp \left\{ \frac{2 \theta_{k-1}}{\sigma^2} x \right\} 
     + \frac{h z_k}{\theta_{k-1}} \exp \left\{ \frac{2 \theta_{k-1}}{\sigma^2} z_k \right\} \\
& + \frac{h \sigma^2}{2\theta_{k-1}^2} \left( \exp \left\{ \frac{2 \theta_{k-1}}{\sigma^2}x \right\}
     - \exp \left\{ \frac{2 \theta_{k-1}}{\sigma^2} z_k \right\} \right).
\end{align*}

Further rearranging the terms gives
\begin{align*}
v(x) \exp \left\{ \frac{2 \theta_{k-1}}{\sigma^2} x \right\} 
     =& \exp \left\{ \frac{2 \theta_{k-1}}{\sigma^2} z_k \right\} 
     \left( p - c_k - \frac{\beta + p\theta_{k-1} - c(\theta_{k-1}) }{\theta_{k-1}} + \frac{h z_k}{\theta_{k-1}}
     - \frac{h \sigma^2}{2 \theta_{k-1}^2} \right) \\
&+ \exp \left\{ \frac{2 \theta_{k-1}}{\sigma^2} x \right\} \left(
      \frac{\beta + p\theta_{k-1} - c(\theta_{k-1}) }{\theta_{k-1}} - \frac{hx}{\theta_{k-1}} +\frac{h \sigma^2}{ 2 \theta_{k-1}^2} \right), 
	\end{align*}		
		which then yield the desired result, completing the proof.
\begin{align*}
v(x) =&  \exp \left\{ - \frac{2 \theta_{k-1}}{\sigma^2} (x - z_k) \right\} 
     \left( p - c_k - \frac{\beta + p\theta_{k-1} - c(\theta_{k-1}) }{\theta_{k-1}} + \frac{h z_k}{\theta_{k-1}}
     - \frac{h \sigma^2}{2 \theta_{k-1}^2} \right) \\
&+  \left(\frac{\beta + p\theta_{k-1} - c(\theta_{k-1}) }{\theta_{k-1}} - \frac{hx}{\theta_{k-1}} + \frac{h \sigma^2}{ 2 \theta_{k-1}^2} \right).
\end{align*}
\end{proof}

%%%%%%%%%%%%%%%%%%%%%%%%%%%%%%%%%%
% A NUMERICAL EXAMPLE
%%%%%%%%%%%%%%%%%%%%%%%%%%%%%%%%%%
\section{A Numerical Example} \label{sec:num}

We compare the performance of the dynamic policy with benchmark static policies using a numerical example. In particular, we consider static policies that set $\theta(t) = \theta <0$ for $ t \ge 0$. Under a static policy, the queue length process $\{ Z(t), t \ge0\}$ is a $( \theta, \sigma^2)$ reflected Brownian motion. As such, the following holds (see \citealt{harrison-2013}):
\begin{align*}
& \frac{1}{t} \mathbb{E}[Z(t)] \rightarrow \frac{\sigma^2}{2 | \theta|} 
   \mbox{ as } t \rightarrow \infty, \\
& \frac{1}{t} \mathbb{E}[L(t)] \rightarrow  |\theta|
   \mbox{ as } t \rightarrow \infty.
\end{align*}
Then letting $\beta(\theta)$ denote the long-run average cost of the static policy for $\theta < 0$, we have that
\begin{align}
\beta(\theta) = c(\theta) + \frac{h \sigma^2}{2 | \theta |} + p |\theta|,
   \,\, \theta < 0. \label{eqn:static}
\end{align}

For the following numerical example, let $K = 4$, $\theta_0 = -1.5$,   $\theta_1 = -1$,  $\theta_2 = -0.3$,  $\theta_3 = -0.125$, $\theta_4 = 2.5$, $\sigma = 2$, $h = 3$,  $c_1 = 5$,  $c_2 = 8$,  $c_3 = 20$,    $c_4 = 50$, and  $p = 100$. Recall that the $c_i$ values correspond to the cost rate of increasing the level of promotion activities. For example, suppose the promotion activities in increasing order of cost were: (i) sending mass emails, (ii) advertising online, (iii) advertising on television and radio, and (iv) outreach activities by public health officials. The static policy of sending mass emails ($\theta_1$) would incur cost rate $c(\theta_1)$, the static policy of advertising online ($\theta_2$) would incur cost rate $c(\theta_2)$, etc.

\begin{figure}[htbp]
\centering
\includegraphics[scale=0.4]{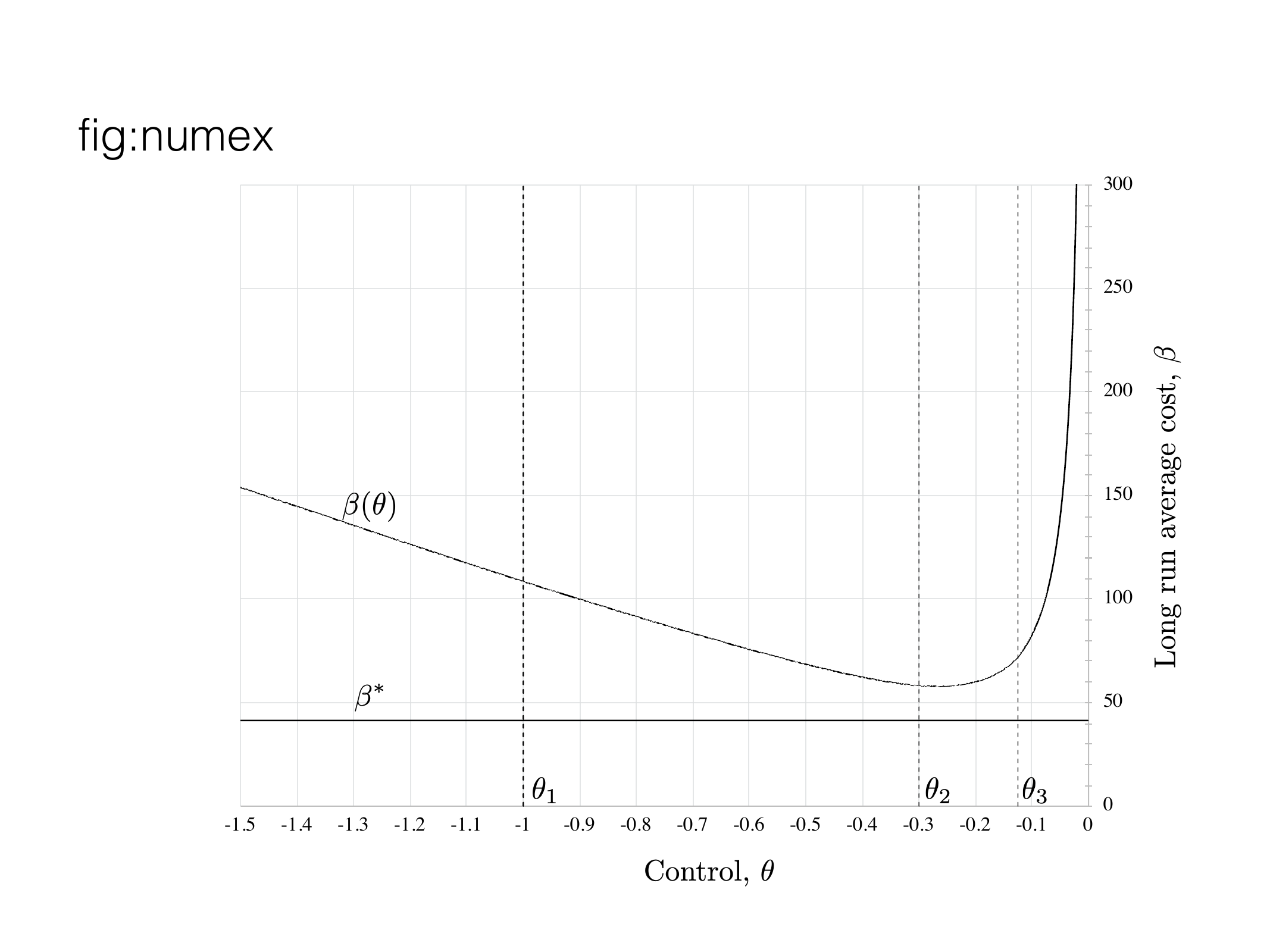}
\caption{Comparing the long run average cost of static policies to the dynamic policy.}
\label{fig:numex}
\end{figure}

We compare the long run average cost of each static policy, $\beta(\theta_i)$, $i = 0, \dots, 4$, given by (\ref{eqn:static}), with the long run average cost $\beta^*$ of the dynamic policy ($\theta^*(z)$ in Theorem~\ref{thm:optpol}). Figure~\ref{fig:numex} shows that the optimal static policy is $\theta_2 = -0.3$, which corresponds to always using the promotion activities of sending mass emails and advertising online. The long run average cost of this static policy is $\beta(-0.3) =58.1$, whereas the long run average cost of the dynamic policy is only $\beta^*= 41.4$. Thus, the cost savings of the dynamic policy over the best static policy is 29\%.\footnote{An alternate ``randomized'' static policy that combines static policy $\theta_2$ with probability 0.85 and static policy $\theta_3$ with probability 0.15 would achieve a long run average cost of 57.9. This is slightly lower than simply consistently using policy $\theta_2$, but the difference is negligible -- the dynamic policy still achieves a cost savings of 28.5\%.}

%%%%%%%%%%%%%%%%%%%%%%%%%%%%%%%%%%
% CONCLUDING REMARKS
%%%%%%%%%%%%%%%%%%%%%%%%%%%%%%%%%%

\section{Concluding Remarks} \label{sec:conc}

Motivated by a public health service context, we studied a diffusion control problem where a system manager can deploy costly promotion activities to increase the rate of participant inflow. We developed a threshold policy and derived explicit queue length thresholds for progressively deploying these promotion activities. Using a numerical study, we showed the potential cost savings of the dynamic policy relative to the optimal static policy can be significant (i.e., in our example, the cost savings was 29\%). Our research makes a methodological contribution by deriving the solution of a drift-rate control problem on an unbounded domain with state costs.

%%%%%%%%%%%%%%%%%%%%%%%%%%
% Bibliography
%%%%%%%%%%%%%%%%%%%%%%%%%%
\newpage

%\bibliographystyle{chicagoa}
%\bibliography{./vol_bib}

%%%%%%%%%%%%%%%%%%%%%%%%%%
% Appendix
%%%%%%%%%%%%%%%%%%%%%%%%%%
\appendix

\section{Proofs} \label{app:proofs}

\paragraph{Proof of Proposition~\ref{prop:equivform}.}
Let $\delta(\cdot)$ be a feasible policy for formulation (\ref{eqn:appox-obj2})-(\ref{eqn:approx-cons2.5}) and set $\theta(t) = \sum_{k=1}^{K} \mu_k \delta_k(t)$ for $t \ge 0$. Clearly, $\theta(t) \in A$ and is feasible for $c(\theta(t)) \le \sum_{k=1}^{K} c_k \mu_k \delta_k(t)$ for $t \ge 0$. Similarly, given a feasible policy $\theta(\cdot)$ for formulation (\ref{eqn:appox-obj3})-(\ref{eqn:approx-cons3.4}), we set
\begin{align}
\delta_k(t) = \min \Bigg\{ 1, \frac{[\theta(t) - \sum_{l=1}^{k-1} \mu_l]^+}{\mu_k} \Bigg\}, \,\, k = 1, \dots, K.
\end{align}
Clearly, $\delta_k(t) \in [0, 1]$ for all $t \ge 0$. Thus, $\delta(\cdot)$ is feasible for formulation (\ref{eqn:appox-obj2})-(\ref{eqn:approx-cons2.5}). Also, it is easy to see from Equation~(\ref{eqn:cx}) that $\sum_{k=1}^{K} c_k \mu_k \delta_k(t) = c(\theta(t))$ for $t \ge 0$. These imply that the two formulations have the same long-run average cost.
$\blacksquare$

%%%%%%%%%%%%%%%%%%%%%%%%%%%%%%
%%%%%%%%%%%%%%%%%%%%%%%%%%%%%%

\paragraph{Proof of Lemma~\ref{lem:vbeta}.}
First, we prove that $v_{\beta_2}(x) > v_{\beta_1}(x)$ for $x>0$ and $\beta_2 > \beta_1 > \betainf$. Fix $\beta_2 > \beta_1 > \betainf$. Define $\tilde{\phi}(y) = \phi(y) - \theta_k y$ and note that $\tilde{\phi}$ is a decreasing function. Then substituting $\phi(y)  = \tilde{\phi}(y) + \theta_k y$ in Equation~(\ref{eqn:ivp1}) and rearranging the terms, we arrive at the following: For $i=1,2$,
\begin{align}
v_{\beta_i}'(x) + \frac{2 \theta_K}{\sigma^2} v_{\beta_i}(x)
     = \frac{2 \beta_i}{\sigma^2} 
     + \frac{2}{\sigma^2} \tilde{\phi}(p - v_{\beta_i}(x))
     + \frac{2 \theta_K p}{\sigma^2}
     - \frac{2hx}{\sigma^2}. \label{eqn:lemvbeta1}
\end{align}

We argue by contradiction. Suppose $v_{\beta_1}(x) > v_{\beta_2}(x)$ for some $x>0$. Let $x^* = \inf \{ x\ge0 : v_{\beta_1}(x) \ge v_{\beta_2}(x) \}$. If $x^* > 0$, then by continuity of $v_{\beta_i}(\cdot)$, we conclude that
\begin{align}
v_{\beta_1}(x^*) = v_{\beta_2}(x^*) \quad \mbox{and} \quad 
     v_{\beta_1}(x) < v_{\beta_2}(x) \quad \mbox{on } [0, x^*). \label{eqn:lemvbeta2}
\end{align}
Multiplying both sides of Equation~(\ref{eqn:lemvbeta1}) with $\exp\{2 \theta_K x/\sigma^2\}$ and integrating over $[0, x^*]$ give the following:
\begin{align}
&\exp \left\{ \frac{2 \theta_K}{\sigma^2} x^* \right\}v_{\beta_i}(x^*)  =
     \int_0^{x^*}  \frac{2 \beta_i}{\sigma^2} \exp \left\{ \frac{2 \theta_K}{\sigma^2} x \right\} dx
     + \frac{2}{\sigma^2 } \int_0^{x^*} \tilde{\phi}( p - v_{\beta_i}(x) ) 
     \exp \left\{ \frac{2 \theta_K}{\sigma^2} x \right\} dx \nonumber \\
     &\quad\quad + \frac{2 \theta_K p}{\sigma^2}  \int_0^{x^*} 
     \exp \left\{ \frac{2 \theta_K p}{\sigma^2} x \right\} dx
     - \frac{2}{\sigma^2 } \int_0^{x^*} \exp \left\{ \frac{2 \theta_K x}{\sigma^2} x \right\} h x \,dx, 
     \,\,\, i=1,2 \label{eqn:lemvbeta3}
\end{align}

Considering Equation~(\ref{eqn:lemvbeta3}) for $\beta_1$ and $\beta_2$ and taking the difference give
\begin{align}
0 &=  \int_0^{x^*}  \frac{2 (\beta_2 - \beta_1)}{\sigma^2} \exp \left\{ \frac{2 \theta_K}{\sigma^2} x \right\} dx
     \nonumber \\
     &+ \frac{2}{\sigma^2} \int_0^{x^*} [ \tilde{\phi}( p - v_{\beta_2}(x) ) - \tilde{\phi}( p - v_{\beta_1}(x) )]
     \exp \left\{ \frac{2 \theta_K}{\sigma^2} x \right\} dx > 0, \label{eqn:lemvbeta4}
\end{align}
where the inequality follows from Equation~(\ref{eqn:lemvbeta3}) and the monotonicity of $\tilde{\phi}$. The contradiction reached in Equation~(\ref{eqn:lemvbeta4}) show $v_{\beta_2}(x) > v_{\beta_1}(x)$ when $x > 0$.

Next, we consider the case of $x^*=0$. If $x^* = 0$, then there exists a sequence $\{x_n\}$ such that $x_n \downarrow 0$ as $n \rightarrow \infty$ and $v_{\beta_1}(x_n) \ge v_{\beta_2}(x_n)$. In particular,
\begin{align}
\frac{v_{\beta_1}(x_n)}{x_n} \ge \frac{v_{\beta_2}(x_n)}{x_n} \quad \mbox{for } n\ge1. \label{eqn:lemvbeta5}
\end{align}
Because $v_{\beta_1}(0) = v_{\beta_2}(0)$, taking the limit in Equation~(\ref{eqn:lemvbeta5}) as $n \rightarrow \infty$ gives $v_{\beta_2}'(0) \le v_{\beta_1}'(0)$. Combining this with Equation~(\ref{eqn:ivp1}) gives
\begin{align}
\beta_2 + \phi(p) - h(0) \le \beta_1 + \phi(p) - h(0), \nonumber
\end{align}
or, $\beta_1 \le \beta_2$, which is a contradiction, proving $v_{\beta_2}(x) > v_{\beta_2}(x)$ for $x>0$.

Second, we prove that $v_{\beta}(x)$ is continuous in $\beta$ on $(\beta, \infty)$. To this end, fix $\beta_2 > \beta_1 > \betainf$ and note from Equation~(\ref{eqn:ivp1}) that
\begin{align}
\frac{1}{2} \sigma^2 v_{\beta_i}'(x) = \beta_i - hx - \phi(p - v_{\beta_i}(x)), \,\,\, x \ge 0, i=1,2. \nonumber
\end{align}
Integrating both sides of this on $[0, y]$, we have that
\begin{align}
v_{\beta_2}(y) &= \frac{2}{\sigma^2} \beta_2 y - \frac{h y^2}{\sigma^2}
     -\frac{2}{\sigma^2} \int_{0}^{y} \phi(p - v_{\beta_2}(s)) ds, \nonumber \\
v_{\beta_1}(y) &= \frac{2}{\sigma^2} \beta_1 y - \frac{h y^2}{\sigma^2}
     -\frac{2}{\sigma^2} \int_{0}^{y} \phi(p - v_{\beta_1}(s)) ds. \nonumber
\end{align}
Taking the difference gives
\begin{align}
v_{\beta_2}(y) - v_{\beta_1}(y) = \frac{2}{\sigma^2} (\beta_2 - \beta_1) y 
     - \frac{2}{\sigma^2} \int_{0}^{y} [\phi(p - v_{\beta_2}(s)) - \phi(p - v_{\beta_1}(s)) ] ds. \nonumber
\end{align}
Take the absolute value of both sides and using the Lipschitz continuity of $\phi$ (see Lemma~\ref{lem:philip}), we conclude that
\begin{align}
|v_{\beta_2}(y) - v_{\beta_1}(y)| \le \frac{2}{\sigma^2} |\beta_2 - \beta_1| y 
     + \frac{2L}{\sigma^2} \int_{0}^{y} |v_{\beta_2}(s) - v_{\beta_1}(s)| ds. \nonumber
\end{align}
Note that letting $F(y) = v_{\beta_2}(y) - v_{\beta_1}(y)$, we have for $y \in [0, x]$ that
\begin{align}
|F(y)| \le \frac{2}{\sigma^2} |\beta_2 - \beta_1| x 
     + \int_{0}^{y} F(s)ds. \nonumber
\end{align}

Thus, by Gronwall's inequality (Lemma~\ref{lem:gronwall}), we write
\begin{align}
|v_{\beta_2}(y) - v_{\beta_1}(y)| \le \frac{2x}{\sigma^2} (\beta_2 - \beta_1) 
     +\exp \left\{ \frac{2L}{\sigma^2} y \right\}, \,\, \forall y \in [0,x]. \nonumber
\end{align}
In particular, $v_{\beta}(x)$ is continuous in $\beta$.
$\blacksquare$

\paragraph{Proof of Lemma~\ref{lem:beta}.}
We proceed in several steps. First, we prove that for all $\beta > \betainf$, $v_{\beta}(\cdot)$ strictly increases its maximum. Suppose not. Then by continuity of $v_{\beta}$ and its derivative, there exists $x_2 > x_1 > 0$ such that 
\begin{align}
0 = v(x_1) &\le v'(x_2), \nonumber\\
v(x_1) &= v(x_2). \nonumber
\end{align}
Then we write by Equation~(\ref{eqn:ivp1}) that 
\begin{align}
\beta &= \frac{1}{2} \sigma^2 v_{\beta}'(x_1) + hx_1 - \phi(p - v_{\beta}(x_1)), \label{eqn:lembeta1} \\
\beta &= \frac{1}{2} \sigma^2 v_{\beta}'(x_2) + hx_2 - \phi(p - v_{\beta}(x_2)). \label{eqn:lembeta2}
\end{align}
Subtracting (\ref{eqn:lembeta1}) from (\ref{eqn:lembeta2}) yields:
\begin{align*}
0 = \frac{1}{2} \sigma^2 v_{\beta}'(x_2) + h(x_2 - x_1) > 0,
\end{align*}
which is a contradiction.

Second, we show that $v_{\beta}$ is either strictly increasing or it is first increasing then decreasing. Because $v_{\beta}$ strictly increases to its maximum, if it never reaches to its maximum, then we are done. Otherwise, let $x_0 < \infty$ denote the maximizer of $v_{\beta}(x)$. It suffices to show that $v_{\beta}(x)$ is decreasing on $(x_0, \infty)$. Suppose not. Then there exists $x_1$, $x_2$ such that
\begin{align*}
& x_0 < x_1 < x_2, \\
& v_{\beta}'(x_1) < 0 < v_{\beta}'(x_2), \\
& v_{\beta}(x_1) = v_{\beta}(x_2).
\end{align*}
It also follows from Equation~(\ref{eqn:ivp1}) that 
\begin{align}
\beta &= \frac{1}{2} \sigma^2 v_{\beta}'(x_1) + hx_1 - \phi(p - v_{\beta}(x_1)), \label{eqn:lembeta3} \\
\beta &= \frac{1}{2} \sigma^2 v_{\beta}'(x_2) + hx_2 - \phi(p - v_{\beta}(x_2)). \label{eqn:lembeta4}
\end{align}
Subtracting (\ref{eqn:lembeta3}) from (\ref{eqn:lembeta4}) yields:
\begin{align*}
0 = \frac{1}{2} \sigma^2 (v_{\beta}'(x_2) -v_{\beta}'(x_1))+ h(x_2 - x_1) > 0,
\end{align*}
which is a contradiction.

Third, we show that for $\beta > \betainf$, if $v_{\beta}$ is increasing, then $\lim_{x \rightarrow \infty} v_{\beta}(x) =\infty$. Suppose not, i.e., $v_{\beta}$ is bounded. Then $v_{\beta}'(x) \rightarrow 0$ as $x \rightarrow \infty$. Recall from Equation~(\ref{eqn:ivp1}) that 
\begin{align*}
\frac{1}{2} \sigma^2 v_{\beta}'(x) = \beta - hx + \phi(p - v_{\beta}(x)), 
\end{align*}
which implies that $\phi(p - v_{\beta}(x)) - hx \rightarrow \beta$ for $x \rightarrow \infty$. This is possible only if $\phi(p - v_{\beta}(x)) = \theta_0 (p - v_{\beta}(x))$ for $x$ large and that $\theta_0 p - \theta_0 v(x) - hx \rightarrow p$ as $x \rightarrow \infty$. That is,
\begin{align*}
v(x) \sim \frac{h}{|\theta_0|}x \quad \mbox{as } x \rightarrow \infty,
\end{align*}
which contradicts that $v_{\beta}$ is bounded.

So far, we proved that for each $\beta > \betainf$, it belongs to either $\mathcal{I}$ or $\mathcal{N}$. Thus, $(\betainf, \infty) = \mathcal{I} \cup \mathcal{N}$. Clearly $\mathcal{I} \cap \mathcal{N} = \emptyset$.

To conclude the proof, it remains to show that if $\beta_1 \in \mathcal{I}$, then $\beta_2 \in \mathcal{I}$ for all $\beta_2 > \beta_1$. Let $\beta_1 \in \mathcal{I}$ and suppose that there exists $\beta_2 > \beta_1$ such that $\beta_2 \notin \mathcal{I}$. That is, $v_{\beta_2}$ is first increasing, then decreasing. Then because $v_{\beta_1}(x) \rightarrow \infty$ as $x \rightarrow \infty$, there exists $x_0$ such that $v_{\beta_1}(x_0) > v_{\beta_2}(x_0)$, which contradicts that $v_{\beta}(x)$ is increasing in $\beta$ (see Lemma~\ref{lem:vbeta}).
$\blacksquare$

\paragraph{Proof of Lemma~\ref{lem:INnotempty}.}
We first prove that $\mathcal{I} \not= \emptyset$ and $\left( p | \theta_0 | + \frac{h \sigma^2 }{2 | \theta_0| }, \infty \right) \subset \mathcal{I}$. That is, for $\beta > p | \theta_0 | + \frac{h \sigma^2 }{2 | \theta_0| }$, $v_{\beta}(x)$ is strictly increasing in $x$ and that $v_{\beta}(x) \rightarrow \infty$ as $ x \rightarrow \infty$. Recall from Equation (\ref{eqn:ivp1}) that
\begin{align}
v'_{\beta}(x) &= \frac{2 \beta}{\sigma^2} + \frac{2}{\sigma^2} \phi ( p - v_{\beta}(x) ) 
     - \frac{2h}{\sigma^2} x, \,\,\, x \ge 0 \label{eqn:INnotempty1} \\
v_{\beta}(0) &= 0. \label{eqn:INnotempty2}
\end{align}
Also note that $\phi(y) \ge \theta_0 y$ for all $y$. Thus,
\begin{align}
v'_{\beta}(x) & \geq \frac{2 \beta}{\sigma^2} + \frac{2 \theta_0}{\sigma^2} ( p - v_{\beta}(x) ) 
     - \frac{2h}{\sigma^2} x, \,\,\, x \ge 0. \label{eqn:INnotempty3} 
\end{align}
Rearranging the terms, we write
\begin{align}
v'_{\beta}(x) + \frac{2 \theta_0}{\sigma^2} v_{\beta}(x) \ge \frac{2 (\beta + \theta_0 p)}{\sigma^2}  
     - \frac{2h}{\sigma^2} x. \label{eqn:INnotempty4} 
\end{align}
Multiplying both sides by $\exp \left\{ \frac{2 \theta_0}{\sigma^2} x\right\}$ gives the following
\begin{align*}
 \left[ \exp \left\{ \frac{2 \theta_0}{\sigma^2} x\right\} v_{\beta}(x) \right]' \ge 
      \exp \left\{ \frac{2 \theta_0}{\sigma^2} x\right\} 
      \frac{2 (\beta + \theta_0 p)}{\sigma^2} - \frac{2h}{\sigma^2}x  
      \exp \left\{ \frac{2 \theta_0}{\sigma^2} x\right\}.
\end{align*}
Integrating both sides and using Equation~(\ref{eqn:INnotempty2}) give the following:
\begin{align*}
\exp \left\{ \frac{2 \theta_0}{\sigma^2} x\right\} v_{\beta}(x) 
     \ge \frac{2 (\beta + \theta_0 p)}{\sigma^2} \frac{\sigma^2}{2 \theta_0} 
     \exp \left\{ \frac{2 \theta_0}{\sigma^2} z\right\} \Bigg|_0^x 
     - \frac{2h}{\sigma^2} \int_0^x z \exp \left\{ \frac{2 \theta_0}{\sigma^2} z\right\} dz.
\end{align*}
That is,
\begin{align}
v_{\beta}(x) \exp \left\{ \frac{2 \theta_0}{\sigma^2} x\right\}  
     \ge \left( \frac{\beta}{\theta_0} + p \right) 
     \left[ \exp \left\{ \frac{2 \theta_0}{\sigma^2} x\right\} - 1\right]
     - \frac{2h}{\sigma^2 } 
     \int_0^x z \exp \left\{ \frac{2 \theta_0}{\sigma^2} z\right\} dz, \label{eqn:INnotempty5}
\end{align}
where the last term can be integrated by parts as follows:
\begin{align*}
\frac{2h}{\sigma^2} \int_0^x z \exp \left\{ \frac{2 \theta_0}{\sigma^2} z\right\} dz 
     &= \frac{2h}{\sigma^2} \frac{\sigma^2}{2 \theta_0} z 
     \exp \left\{ \frac{2 \theta_0}{\sigma^2} z\right\} \Bigg|_0^x
     - \frac{2h}{\sigma^2 }
      \int_0^x \frac{\sigma^2}{2 \theta_0} \exp \left\{ \frac{2 \theta_0}{\sigma^2} z\right\} dz \\
&= \frac{h}{\theta_0} z \exp \left\{ \frac{2 \theta_0}{\sigma^2} z\right\} \Bigg|_0^x
     - \frac{h}{\theta_0} \int_0^x  \exp \left\{ \frac{2 \theta_0}{\sigma^2} z\right\} dz \\
&= \frac{h}{\theta_0} x \exp \left\{ \frac{2 \theta_0}{\sigma^2} x \right\}
     - \frac{h}{\theta_0}  \frac{\sigma^2}{2 \theta_0} \exp \left\{ \frac{2 \theta_0}{\sigma^2} x \right\}
     \Bigg|_0^x \\
&= \frac{h}{\theta_0} x \exp \left\{ \frac{2 \theta_0}{\sigma^2} x \right\}
     - \frac{h \sigma^2}{2 \theta_0^2} \exp \left\{ \frac{2 \theta_0}{\sigma^2} x \right\}  
     + \frac{h \sigma^2}{2 \theta_0^2}
\end{align*}
Substituting this into Equation~(\ref{eqn:INnotempty5}) gives the following:
\begin{align*}
v_{\beta}(x) \exp \left\{ \frac{2 \theta_0}{\sigma^2} x\right\} 
     \ge& \left( \frac{\beta}{\theta_0} + p \right)
     \left[ \exp \left\{ \frac{2 \theta_0}{\sigma^2} x\right\} - 1\right]
     - \frac{h}{\theta_0} x \exp \left\{ \frac{2 \theta_0}{\sigma^2} x \right\} \\
&+ \frac{h \sigma^2}{2 \theta_0^2} \exp \left\{ \frac{2 \theta_0}{\sigma^2} x\right\} 
     - \frac{h \sigma^2}{2 \theta_0^2} \\
&= \exp \left\{ \frac{2 \theta_0}{\sigma^2} x\right\}
     \left[ \frac{\beta}{\theta_0} + p - \frac{h}{\theta_0}x 
     + \frac{h \sigma^2}{2 \theta_0^2} \right]
     - \left( \frac{\beta}{\theta_0} + p + \frac{h \sigma^2}{2 \theta_0^2} \right).
\end{align*}
Therefore, we conclude that
\begin{align}
v_{\beta}(x) \ge \left( -\frac{\beta}{\theta_0} - p - \frac{h \sigma^2}{2 \theta_0^2} \right)
     \exp \left\{ \frac{-2 \theta_0}{\sigma^2} x\right\}
     + \left( \frac{\beta}{\theta_0} + p -\frac{h}{\theta_0}x + \frac{h \sigma^2}{2 \theta_0^2} \right).
     \label{eqn:INnotempty6}
\end{align}
Thus, for $\beta > p | \theta_0| + \frac{h \sigma^2}{2 | \theta_0|}$, we have that $v_{\beta}(x) \rightarrow \infty$ as $ x \rightarrow \infty$.
Moreover, note from Equation~(\ref{eqn:INnotempty4}) that
\begin{align}
v_{\beta}'(x) \ge \frac{2( \beta + \theta_0 p)}{\sigma^2} - \frac{2h}{\sigma^2} x
     - \frac{2 \theta_0}{\sigma^2} v_{\beta}(x). \label{eqn:INnotempty7}
\end{align}
Because $\theta_0 < 0$ by assumption, substituting~(\ref{eqn:INnotempty6}) into~(\ref{eqn:INnotempty7}) gives
\begin{align*}
v_{\beta}'(x) \ge - \frac{2h}{\theta_0} + \frac{2}{\sigma^2}
     \left( \beta - p |\theta_0| - \frac{h \sigma^2}{2 | \theta_0|} \right)
     \exp \left\{ -\frac{2 \theta_0 x }{\sigma^2} \right\} > 0,
\end{align*}
where the last inequality follows because $\theta_0 < 0$ and $\beta > p |\theta_0| + \frac{h \sigma^2}{2 |\theta_0|}$. This proves that $v_{\beta}(x)$ is strictly increasing for $\beta > p |\theta_0| + \frac{h \sigma^2}{2 |\theta_0|}$, proving the claim.

Next, we prove $\mathcal{N} \not= \emptyset$ by contradiction. As a preliminary, define $q = \inf \{ y \in [0,p] : \phi(y) = -\betainf \}$ and note that $\phi(0) = 0$, $\phi$ is decreasing $(0, q)$ and increasing on $(q,p)$. We have the following three cases to consider:

\noindent \emph{Case 1.} $q=p$. In this case, $\phi$ is decreasing on $(0, p]$ and $\theta_k < 0$.

\noindent \emph{Case 2.} $q<p$ and $\phi(p) \ge 0$. In this case, $\phi(p) = \max_{y \in [0,p]} \phi(y)$ and $\theta_k > 0$.

\noindent \emph{Case 3.} $q<p$ and $\phi(p) < 0$. In this case, $\phi(0) =0= \max_{y \in [0,p]} \phi(y)$ and $\theta_k \ge 0$.

Suppose $\mathcal{N} = \emptyset$, i.e., $v_{\beta}(x)$ increases strictly to $\infty$ for all $\beta > \betainf$ by Lemma~\ref{lem:beta}. In particular, its inverse $v_{\beta}^{-1}$ is well defined and strictly increasing. We will illustrate a $\beta \in ( \betainf, \betainf + 1)$ in each case that contradict this. To this end, for $\beta \in ( \betainf, \betainf + 1)$, let
\begin{align*}
\bar{x}(\beta) &= \inf \{ x \ge 0 : v_{\beta}(x) = p \} = v_{\beta}^{-1}(p), \\
x_0(\beta) &= \inf \{ x \ge 0 : p - v_{\beta}(x) = q \} = v_{\beta}^{-1}(p-q), \\
\hat{x}(\beta) &= \inf \{ x \ge 0 : v_{\beta}(x) = c_k \} = v_{\beta}^{-1}(c_k), \\
\tilde{x}(\beta) &= \inf \left\{ x \ge 0 : v_{\beta}(x) = \frac{p - c_k}{2} \right\} 
     = v_{\beta}^{-1} \left(\frac{p - c_k}{2} \right).
\end{align*}
Next, we consider each of the above cases.

\noindent \emph{Case 1.} In this case, $x_0(\beta) = 0$ because $v_{\beta}(0)=0$ and $p - v_{\beta}(0) = p-q$. Moreover, for $x \in [0, \hat{x}(\beta))$, we have that
\begin{align}
\phi(p - v_{\beta}(x)) = \phi(p) - \theta_k v_{\beta}(x) > \phi(p). \label{eqn:INnotempty8}
\end{align}
Recall from Equation~(\ref{eqn:ivp1}) that
\begin{align}
v_{\beta}'(x) &= \frac{2 \beta}{\sigma^2} + \frac{2}{\sigma^2} \phi( p - v_{\beta}(x))
     - \frac{2h}{\sigma^2}x, \,\,\, x \ge 0 \label{eqn:INnotempty9} \\
v_{\beta}(0) &= 0. \label{eqn:INnotempty10}
\end{align}
For $\varepsilon \in (0,1)$, let $\beta = \betainf + \varepsilon$. Then substituting~(\ref{eqn:INnotempty8}) into~(\ref{eqn:INnotempty9}) gives
\begin{align*}
v_{\beta}'(x) &= \frac{2 \beta}{\sigma^2} + \frac{2 \varepsilon}{\sigma^2}
     + \frac{2 \phi(p)}{\sigma^2} - \frac{2 \theta_k}{\sigma^2} v_{\beta}(x)
     - \frac{2h}{\sigma^2} x \\
&= \frac{2 \varepsilon}{\sigma^2} - \frac{2 \theta_k}{\sigma^2} v_{\beta}(x)
     - \frac{2h}{\sigma^2} x,
\end{align*}
because $\phi(p) = -\betainf$. Rearranging the terms further gives
\begin{align*}
v_{\beta}' + \frac{2 \theta_k}{\sigma^2} v_{\beta}(x) 
     = \frac{2 \varepsilon}{\sigma^2} - \frac{2h}{\sigma^2}x, \,\,\, x \in [0, \hat{x}(\beta)].
\end{align*}
Multiplying both sides by $\exp \left\{ \frac{2 \theta_k}{\sigma^2}x \right\}$ gives
\begin{align*}
\left[ \exp\left\{  \frac{2 \theta_k }{\sigma^2} x \right\} v_{\beta}(x) \right]'
     = \frac{2 \varepsilon}{\sigma^2} \exp\left\{  \frac{2 \theta_k}{\sigma^2} x \right\}
     - \frac{2h}{\sigma^2} x \exp\left\{  \frac{2 \theta_k}{\sigma^2} x \right\}.
\end{align*}
Integrating both sides from 0 to $x \in (0, \hat{x}(\beta))$ gives:
\begin{align}
\exp\left\{  \frac{2 \theta_k }{\sigma^2} x \right\} v_{\beta}(x) 
     &= \frac{2 \varepsilon}{\sigma^2} \frac{\sigma^2}{2 \theta_k} 
     \exp\left\{  \frac{2 \theta_k }{\sigma^2} z \right\} \Bigg|_0^x
     - \int_o^x \frac{2h}{\sigma^2} z \exp \left\{ \frac{2 \theta_k}{\sigma^2} z \right\} dz \nonumber \\
&= \frac{\varepsilon}{\theta_k} \exp \left\{ \frac{2 \theta_k}{\sigma^2} x \right\} 
     - \frac{\varepsilon}{\theta_k} - \int_0^x \frac{2h}{\sigma^2} z
     \exp \left\{ \frac{2 \theta_k}{\sigma^2} z \right\} dz. \label{eqn:INnotempty11}
\end{align}
As done earlier in the proof, the last term on the right-hand side of~(\ref{eqn:INnotempty11}) can be integrated by parts to give the following:
\begin{align*}
\frac{2h}{\sigma^2} \int_0^x z \exp \left\{ \frac{2 \theta_k}{\sigma^2} z \right\} dz 
     = \frac{h}{\theta_k} x \exp \left\{ \frac{2 \theta_k}{\sigma^2} x \right\}
     - \frac{h \sigma^2}{2\theta_k^2} x \exp \left\{ \frac{2 \theta_k}{\sigma^2} x \right\}
     + \frac{h \sigma^2}{2\theta_k^2}.
\end{align*}
Substituting this into Equation~(\ref{eqn:INnotempty11}) gives
\begin{align*}
\exp \left\{ \frac{2 \theta_k}{\sigma^2} x \right\} v_{\beta}(x)
     = \left[  \frac{\varepsilon}{\theta_k} - \frac{h}{\theta_k}x 
     + \frac{h \sigma^2}{2 \theta_k^2}\right] \exp \left\{ \frac{2 \theta_k}{\sigma^2} x \right\}
     - \left( \frac{\varepsilon}{\theta_k} +  \frac{h \sigma^2}{2 \theta_k^2} \right).
\end{align*}
Thus, we have that
\begin{align*}
 v_{\beta}(x) = \left[  \frac{\varepsilon}{\theta_k} - \frac{h}{\theta_k}x 
     + \frac{h \sigma^2}{2 \theta_k^2}\right] 
     - \left(\frac{h \sigma^2}{2 \theta_k^2} + \frac{\varepsilon}{\theta_k}  \right)
     \exp \left\{ -\frac{2 \theta_k}{\sigma^2} x \right\}
\end{align*}
Differentiating both sides gives
\begin{align*}
 v_{\beta}'(x) = -\frac{h}{\theta_k} + \frac{h}{\theta_k} \exp \left\{ -\frac{2 \theta_k}{\sigma^2} x \right\}
      + \frac{2 \varepsilon}{\sigma^2} \exp \left\{ -\frac{2 \theta_k}{\sigma^2} x \right\}.
\end{align*}
Rearranging the terms, we can rewrite this as follows:
\begin{align}
 v_{\beta}'(x) = \exp \left\{ -\frac{2 \theta_k}{\sigma^2} x \right\}
     \left[ \frac{h}{\theta_k} \left( 1 - \exp \left\{ \frac{2 \theta_k}{\sigma^2} x \right\} \right)
     + \frac{2 \varepsilon}{\sigma^2}\right]. \label{eqn:INnotempty12}
\end{align}

Next, we verify that for $\beta \in (\betainf, \betainf + 1)$,
\begin{align}
\tilde{x}(\beta) \ge \frac{(p-c_k) \sigma^2}{4(\betainf+1)}. \label{eqn:INnotempty13}
\end{align}
Recall that $\hat{x}(\beta) \ge \tilde{x}(\beta)$. Note by definition of $\tilde{x}(\beta)$ that
\begin{align}
v_{\beta}(\tilde{x}(\beta)) = \frac{p-c_k}{2}. \label{eqn:INnotempty14}
\end{align}
Also note from Equation~(\ref{eqn:ivp1}) that
\begin{align}
v_{\beta}'(x) &= \frac{2 \beta}{\sigma^2} + \frac{2}{\sigma^2}
     \phi (p - v_{\beta}(x)) - \frac{2h}{\sigma^2} x, \,\,\, x \in (0, \hat{x}(\beta)) \nonumber\\
     &\le \frac{2 \beta}{\sigma^2} \nonumber \\
     &\le \frac{2(\betainf + 1)}{\sigma^2},  \label{eqn:INnotempty15}
\end{align}
where the first inequality follows because $\phi(p - v_{\beta}(x)) < 0$ for $x \in (0, \hat{x}(\beta))$, and the second inequality follows because we restrict attention to $\beta \in (\betainf, \betainf + 1)$. Integrating both sides of Equation~(\ref{eqn:INnotempty15}) on $(0, \hat{x}(\beta))$ gives
\begin{align*}
\frac{p - c_k}{2} = v_{\beta}(\tilde{x}(\beta)) \le \frac{2 (\betainf+1)}{\sigma^2}
     \tilde{x}(\beta),
\end{align*}
which follows from~(\ref{eqn:INnotempty14})-(\ref{eqn:INnotempty15}) and establishes~(\ref{eqn:INnotempty13}).

Consider the derivative of the second term of the product on the right-hand side of Equation~(\ref{eqn:INnotempty12}):
\begin{align*}
\frac{d}{dx} \left[ \frac{h}{\theta_k} \left( 1 - \exp \left\{ \frac{2 \theta_k}{\sigma^2} x \right\} \right)
     + \frac{2 \varepsilon}{\sigma^2}\right]
     = -\frac{2h}{\sigma^2} \exp \left\{ \frac{2 \theta_k}{\sigma^2} x \right\}  < 0.
\end{align*}
Thus, we conclude from Equation~(\ref{eqn:INnotempty13}) that
\begin{align*}
\frac{h}{\theta_k} \left( 1 - \exp \left\{ \frac{2 \theta_k}{\sigma^2} \tilde{x}(\beta) \right\}\right)
     + \frac{2 \varepsilon}{\sigma^2} 
     < \frac{h}{\theta_k} \left( 1 - \exp \left\{ \frac{ \theta_k (p - c_k)}{2(\betainf +1)} \right\}\right)
     + \frac{2 \varepsilon}{\sigma^2}.
\end{align*}
In particular, combining this with Equation~(\ref{eqn:INnotempty12}) gives
\begin{align*}
v_{\beta}'(\tilde{x}(\beta)) \le \exp \left\{ -\frac{2 \theta_k}{\sigma^2} \tilde{x}(\beta) \right\}
     \left[ \frac{h}{\theta_k}  
     \left( 1 - \exp \left\{ \frac{ \theta_k (p - c_k)}{2(\betainf +1)} \right\}\right)
     + \frac{2 \varepsilon}{\sigma^2} \right].
\end{align*}
Letting $\varepsilon \in \left( 0, \frac{\sigma^2}{4} \frac{h}{| \theta_k|} \left( 1 - \exp \left\{ \frac{ \theta_k (p - c_k)}{2(\betainf +1)} \right\}\right) \wedge 1 \right)$, we conclude that $v_{\beta}'(\tilde{x}(\beta)) < 0$, where $\beta = \betainf + \varepsilon$, which contradicts that $v_{\beta}$ is strictly increasing. 

To conclude the proof, we consider cases~2 and~3 and treat them simultaneously. Recall that $x_0(\beta) = v_{\beta}^{-1}(p-q)$, i.e., $v_{\beta}(x_0(\beta)) = p - q$. Also note from Equation~(\ref{eqn:ivp1}) that
\begin{align*}
v_{\beta}'(x) = \frac{2 \beta}{\sigma^2} + \frac{2}{\sigma^2} \phi( p - v_{\beta}(x))
     - \frac{2h}{\sigma^2}x, \,\,\, x \in (0, \bar{x}(\beta)).
\end{align*}
In particular,
\begin{align}
v_{\beta}'(x) \le \frac{2 \beta}{\sigma^2} + \frac{2}{\sigma^2} \phi(p)^+, \label{eqn:INnotempty16}
\end{align}
where $\phi(p)^+ = \max \{ \phi(p), 0\}$. This follows because $\phi(p)^+ \ge \phi(y)$ for $y \in [0, p]$ in cases 2 and 3. Integrating both sides of~(\ref{eqn:INnotempty16}) on $(0, x_0(\beta))$ gives
\begin{align*}
p-q = v_{\beta} (x_0(\beta)) \le \frac{2}{\sigma^2} (\beta + \phi(p)^+) x_o(\beta),
\end{align*}
from which it follows that
\begin{align}
x_0(p) \ge \frac{\sigma^2}{2} \frac{p-q}{\beta + \phi(p)^+} 
     \ge \frac{\sigma^2}{2}  \frac{p-q}{\betainf +1+ \phi(p)^+}, \label{eqn:INnotempty17}
\end{align}
where the last inequality follows because we restrict attention to $\beta \in (\betainf, \betainf+1)$. 

For $\varepsilon \in (0, 1)$, letting $\beta = \betainf + \varepsilon$, we note from Equation~(\ref{eqn:ivp1}) that
\begin{align}
v_{\beta}'( x_0(\beta)) &= \frac{2(\betainf + \varepsilon)}{\sigma^2}
     + \frac{2}{\sigma^2} \phi( p - v_{\beta}( x_0(\beta))) - \frac{2h}{\sigma^2}  x_0(\beta) \nonumber \\
&= \frac{2 \varepsilon}{\sigma^2} + \frac{2 \beta}{\sigma^2} + \frac{2}{\sigma^2} \phi(q)
     - \frac{2 h}{\sigma^2} x_0(\beta) \nonumber \\
&= \frac{2 \varepsilon}{\sigma^2}  - \frac{2 h}{\sigma^2} x_0(\beta), \label{eqn:INnotempty18}
\end{align}
where the last equality follows because $\phi(q) = -\betainf$. Letting $\varepsilon \in \left( 0, \frac{h \sigma^2}{4} \frac{p-q}{\betainf +1 + \phi(p)^+} \wedge 1 \right)$, we conclude from~(\ref{eqn:INnotempty18}) that for $\beta = \betainf + \varepsilon$, 
\begin{align*}
v_{\beta}'(x_0(\beta)) = \frac{2}{\sigma^2} (\varepsilon - h x_0(\beta))
     \le \frac{2}{\sigma^2} \left( \varepsilon - \frac{\sigma^2 h}{2}
     \frac{p-q}{\betainf + 1 + \phi(p)^+} \right) < 0,
\end{align*}
where the first inequality follows from~(\ref{eqn:INnotempty17}). This contradicts that $v_{\beta}$ is strictly increasing, concluding the proof.
$\blacksquare$

% Proof of Lemma~\ref{lem:propertiesf}.
\paragraph{Proof of Lemma~\ref{lem:propertiesf}.} To establish part i), it suffices to show the following (see \citealt{harrison-2013}):
\begin{align}
\mathbb{E} \left[ \int_0^t (f'(Z(s)))^2 ds \right] < \infty \quad \forall t > 0. \label{eqn:lempropertiesf1}
\end{align}
To verify (\ref{eqn:lempropertiesf1}), recall that $f'(z) = v(z) - p$ and consider the following:
\begin{align}
\mathbb{E} \left[ \int_0^t (v(Z(s)) - p)^2 ds \right] 
     =& \,\mathbb{E} \left[ \int_0^t ( v^2(Z(s)) - 2p v(Z(s)) + p^2) ds \right] \nonumber \\
\le& \mathbb{E} \int_0^t v^2(Z(s)) ds +2 p \mathbb{E} \int_0^t v(Z(s)) ds + p^2. \label{eqn:lempropertiesf2}
\end{align}
Recall that $v(\cdot)$ has linear growth. Thus, there exists exists $K_1, K_2 > 0$ such that
\begin{align}
v(z) \le K_1 + K_2 z, \,\,\, z > 0. \label{eqn:lempropertiesf3}
\end{align}
Substituting this into Equation~(\ref{eqn:lempropertiesf2}) yields
\begin{align*}
\mathbb{E} \left[ \int_0^t ( v(Z(s)) - p)^2 ds \right] 
     \le& \,\, \mathbb{E}  \int_0^t (K_1^2 + 2 K_1 K_2 Z(s) + K_2^2 Z^2(s)) ds \\
&+ 2p \mathbb{E} \int_0^t (K_1 + K_2 Z(s)) ds + p^2.
\end{align*}
It is straightforward to show that the right-hand side is finite, proving part i).

To establish part ii), recall that under an admissible policy $\theta(\cdot)$, $\exists \, \bar{z}>0$ such that $\theta(z) \le \theta_{j^*} < 0$ for $z \ge \bar{z}$. Fix an admissible policy $\theta(\cdot)$ and the associated $\bar{z}$. Also, let $\tilde{B}$ be a $(\theta_{j^*}, \sigma^2)$ Brownian motion, and let $\tilde{Z}(\cdot)$ be the associated reflected Brownian motion. It is straightforward to argue that $\bar{z} + \tilde{Z}(t)$ is stochastically larger than the (scaled) queue-length process $Z(\cdot)$ under $\theta(\cdot)$. Moreover, note that 
\begin{align*}
f(z) = \int_0^z (v(z) - p) ds = \int_0^z v(s) ds - pz.
\end{align*}
In particular, we have
\begin{align*}
|f(z)| \le  \int_0^z |v(s)| ds + pz.
\end{align*}
Because $v(\cdot)$ has linear growth (see Equation~(\ref{eqn:lempropertiesf3})), we conclude that 
\begin{align*}
|f(z)| \le K_1 z + \frac{K_2}{2} z^2 + pz.
\end{align*}

Thus, we have that
\begin{align}
\mathbb{E} |f(Z(t))| &\le (K_1 + p) \mathbb{E} Z(t) + \frac{K_2}{2} \mathbb{E} Z(t)^2 \nonumber\\
&\le (K_1 + p) \mathbb{E} (\bar{z} + \tilde{Z}(t)) 
     + \frac{K_2}{2} \mathbb{E} (\bar{z} + \tilde{Z}(t))^2, \label{eqn:lempropertiesf4}
\end{align}
where the last inequality follows $(\bar{z} + \tilde{Z}(t))$ is stochastically larger than $Z(t)$. 

Also note that $\tilde{Z}(t) \Rightarrow \tilde{Z}^*$ as $t \rightarrow \infty$, where $\tilde{Z}^*$ is an exponential random variable with mean $\sigma^2/2| \theta_{j}|$ \citep{harrison-2013}. Therefore, we deduce from Equation~(\ref{eqn:lempropertiesf4}) that 
\begin{align*}
\overline{\lim}_{t \rightarrow \infty} \mathbb{E} |f(Z(t))| \le (p + K_1) ( \bar{z} + \mathbb{E} Z^*)
     + \frac{K_2}{2} \mathbb{E}( \bar{z} + Z^*)^2.
\end{align*}
Thus, 
\begin{align*}
\lim_{t \rightarrow \infty} \frac{\mathbb{E} |f(Z(t))| }{t} = 0.
\end{align*}
\hfill $\blacksquare$

\end{document}